\documentclass[acmtog]{acmart}

\usepackage{amsmath}
\usepackage{xcolor, soul}
\usepackage{bm}
\usepackage{booktabs} %
\usepackage{float}
\usepackage{afterpage}
\usepackage{pgf}
\usepackage{tabularx}
\usepackage{nicefrac}
\newcommand{\mycomment}[1]{}

\newcommand{\ignore}[1]{}
\usepackage{graphicx,wrapfig}

\newcommand{\tml}[1]{}
\newcommand{\ravi}[1]{}
\newcommand{\sai}[1]{}

\author{Yash Belhe}
\affiliation{%
  \institution{University of California, San Diego}
  \country{USA}
  }
\email{ybelhe@ucsd.edu}

\author{Bing Xu}
\affiliation{%
  \institution{University of California, San Diego}
  \country{USA}
  }
\email{b4xu@ucsd.edu}
  
\author{Sai Praveen Bangaru}
\affiliation{%
  \institution{Massachusetts Institute of Technology}
  \country{USA}
  }
\email{sbangaru@mit.edu}

\author{Ravi Ramamoorthi}
\affiliation{%
  \institution{University of California, San Diego}
  \country{USA}
  }
\email{ravir@ucsd.edu}

\author{Tzu-Mao Li}
\affiliation{%
  \institution{University of California, San Diego}
  \country{USA}
  }
\email{tzli@ucsd.edu}

\definecolor{yash}{rgb}{0.9,0.7,0.7}

\newcommand{\eqn}[1]{Eqn.~#1}
\newcommand{\figref}[1]{Fig.~\ref{#1}}
\newcommand{\secref}[1]{Sec.~\ref{#1}}
\newcommand{\wi}{\bm \omega_i}
\newcommand{\wid}{\bm \omega_{i,d}}
\newcommand{\wis}{\bm \omega_{i,s}}
\newcommand{\wo}{\bm \omega_o}
\newcommand{\wh}{\bm \omega_h}
\newcommand{\whone}{\bm \omega_{h,1}}
\newcommand{\whtwo}{\bm \omega_{h,2}}
\newcommand{\y}{\bm y}
\newcommand{\z}{\bm z}

\begin{document}

\title{Importance Sampling BRDF Derivatives}
\begin{abstract}

We propose a set of techniques to efficiently importance sample the derivatives of several BRDF models.
BRDF importance sampling is a crucial variance reduction technique in forward rendering.
In differentiable rendering, BRDFs are replaced by their differential BRDF counterparts which are real-valued and can have negative values.
This leads to a new source of variance arising from their change in sign.
Real-valued functions cannot be perfectly importance sampled
by a positive-valued PDF and the direct application of BRDF sampling leads to high variance.
Previous attempts at antithetic sampling only addressed the derivative with the roughness parameter of isotropic microfacet BRDFs.  Our work generalizes BRDF derivative sampling to anisotropic microfacet models, mixture BRDFs, Oren-Nayar, Hanrahan-Krueger, among other analytic BRDFs.

Our method first decomposes the real-valued differential BRDF into a sum of single-signed functions, eliminating variance from a change in sign.
Next, we importance sample each of the resulting single-signed functions separately.
The first decomposition, positivization, partitions the real-valued function based on its sign, and is effective at variance reduction when applicable.
However, it requires analytic knowledge of the roots of the differential BRDF, and for it to be analytically integrable too.
Our key insight is that the single-signed functions can have overlapping support, which significantly broadens the ways we can decompose a real-valued function.
Our product and mixture decompositions exploit this property, and they allow us to support several BRDF derivatives that positivization could not handle.
For a wide variety of BRDF derivatives, our method significantly reduces the variance (up to 58x in some cases) at equal computation cost and enables better recovery of spatially varying textures through gradient-descent-based inverse rendering.

\end{abstract}

\begin{teaserfigure}
  \includegraphics[width=\textwidth]{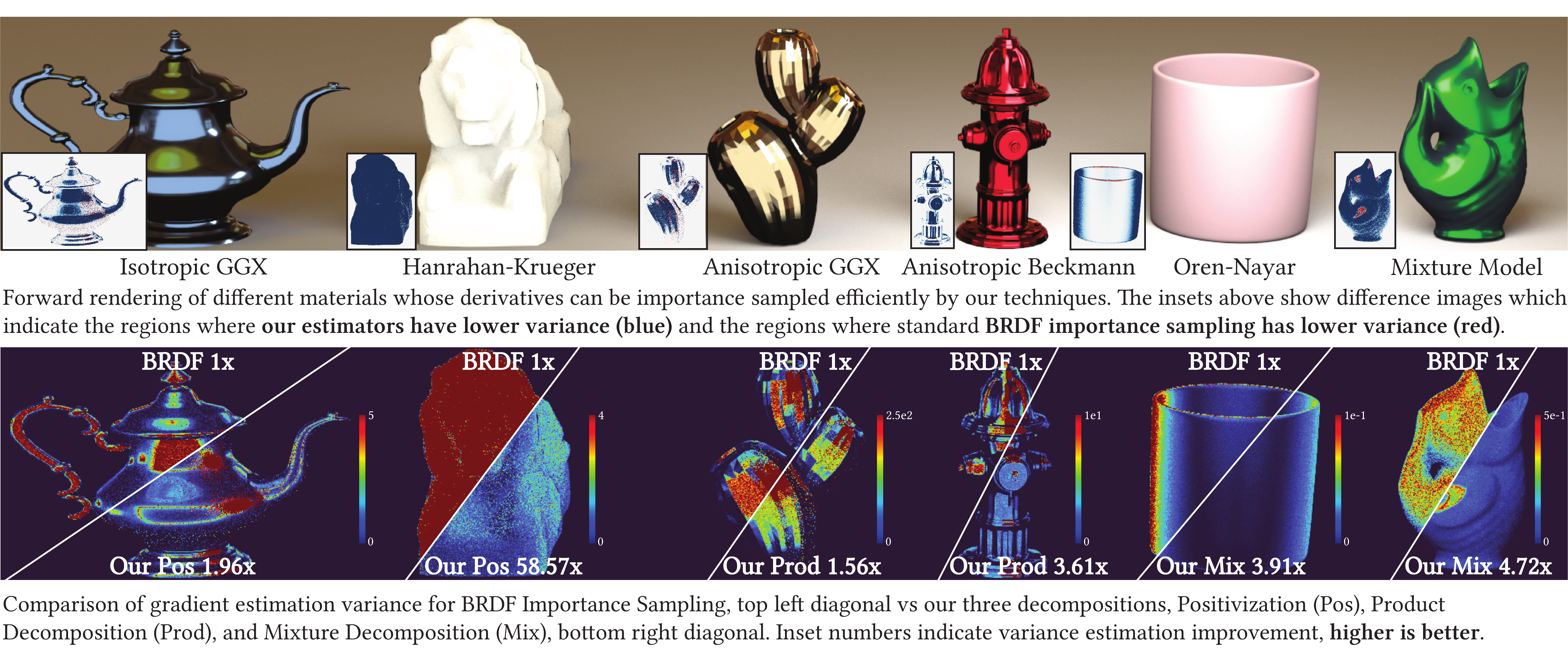}
  \vspace{-0.9cm}
  \caption{\label{fig:teaser}
  We propose a new set of importance sampling techniques for sampling derivatives of BRDFs in rendering, and they achieve significant variance reduction in the estimated derivatives.
  Our techniques work better because they correctly deal with real-valued BRDF derivatives, for which BRDF importance sampling from forward rendering is not well suited.
  Our techniques are general and apply to a wide variety of BRDF derivatives, which was not possible by previous work in differentiable rendering~\cite{Zeltner, Antithetic}, enabling low variance estimates of the derivative at limited sample counts.
  }
\end{teaserfigure}

\maketitle
\section{Introduction}

BRDF importance sampling is an essential variance reduction technique for Monte Carlo forward rendering.
However, there is no simple counterpart for  
differentiable rendering.  
Taking the derivative of a BRDF with respect to one of its parameters transforms it into a real-valued differential BRDF.
The differential BRDF can have a very different shape from the BRDF, and can also be negative valued.
As a result, na\"ively applying BRDF sampling from forward rendering can lead to an estimator with %
a very high variance.
Previous attempts at tackling this problem~\cite{Zeltner} are limited to the roughness derivatives of isotropic GGX (Trowbridge-Reitz) and Beckmann BRDFs, and cannot handle even their anisotropic counterparts. Another method~\cite{Antithetic} was developed primarily for odd functions with symmetric positive and negative lobes, and can produce substantially higher variance when the derivative is close to an even function.
We propose effective importance sampling of derivatives of not only anisotropic GGX and Beckmann BRDFs, but also a wide variety of other analytic BRDF models like Ashikhmin-Shirley, Oren-Nayar, Hanrahan-Krueger, Mixture BRDFs, and ABC models.
\figref{fig:teaser} demonstrates the benefits of our method on several BRDFs compared to BRDF sampling.

Importance sampling a real-valued function leads to unique challenges.
Its variance has two sources, \textit{a) its sign, and b) its shape}.
Our idea is to decompose the function into a sum of single-signed functions (all positive or negative), which we call \textbf{single-signed decompositions}.
Single-signed functions, by definition, have no sign variance.
Importance sampling these functions eliminates their shape variance.

A classical strategy, positivization~\cite{OwenZhou2000}, is a special case of our single-signed decomposition.
It has positive and negative parts with \textit{non-overlapping support}, which in turn requires a) analytic knowledge of the roots and b) analytic integrability of the BRDF derivative up to the roots, which is possible only for certain BRDF derivatives.
To sidestep these issues due to a partition of the domain, we introduce the product and mixture decompositions for which \textit{we allow the positive and negative parts to overlap.}
In fact, we ensure that both the positive and the negative parts have support over the entire hemisphere. This enables analytic integrability and significantly expands upon the set of BRDF derivatives we can handle.
\textit{Our main contributions are three single-signed decompositions and the corresponding importance sampling PDFs of a large set of BRDF derivatives, see Table \ref{table:brdf_list}.}

\textbf{Positivization:} First, we introduce a simple decomposition called positivization
(\secref{sec:positivization}), which \textit{partitions a real-valued function about its roots} into a positive and a negative function.
We show that Zeltner et al.'s~\shortcite{Zeltner} antithetic sampling is a special case of positivization, and positivization provides an explanation of both the correctness and efficiency of their approach.
When applicable, positivization leads to significant variance reduction, for example, the isotropic GGX, Beckmann and Hanrahan-Krueger BRDF derivatives.
However, others like anisotropic GGX, Beckmann, Ashikhmin-Shirley (\secref{sec:positivization_anisotropic}) are not analytically integrable up to their roots, and the derivatives with mixture weights (\secref{subsec:mixture_weights}) do not have analytic roots. Positivization cannot handle these derivatives.
Zeltner et al.'s antithetic sampling inherits these limitations too.

\textbf{Product Decomposition:} Second, we propose a \textit{novel} product decomposition (\secref{sec:product_rule_decomposition}).
Our key observation is that after differentiation, many BRDF derivatives can be decomposed into single-signed functions by \emph{separating the terms that result from the derivative product rule}.
Product decomposition does \emph{not} require knowledge of the roots for the decomposition and only requires the resulting single-signed functions to be analytically integrable.
Product decomposition can importance sample the derivatives of anisotropic GGX, Beckmann, Ashikhmin-Shirley, and more.

\textbf{Mixture Decomposition:} Finally, we introduce mixture decomposition (\secref{subsec:mixture_weights}).
Derivatives of BRDFs with linear combination coefficients, e.g., mixture weights of a layered BRDF, result in real-valued functions whose roots cannot be found analytically in most cases.
Our mixture decomposition exploits the fact that this derivative is the difference between two positive-valued terms.
Separating them results in a single-signed decomposition, and the two terms can then be importance sampled separately.
Mixture decomposition handles the derivatives of Oren-Nayar and mixture weights of \emph{Uber} BRDFs such as the Disney BRDF or Autodesk Standard Surface.

It is likely that several other BRDF derivatives not surveyed in this paper can also be dealt with by one of our three decompositions, and we provide a recipe for handling them in \secref{sec:algorithm}.
We provide a library of importance sampling PDFs for the derivatives of all the BRDF models discussed are above in Table \ref{table:brdf_list}.

\section{Related Work}

Our work connects two areas in rendering research, differentiable rendering and BRDF sampling.
\subsection{Differentiable Rendering}
\noindent\textit{History of derivatives in rendering.} 
Computing derivatives or gradients of light transport has a long
history. Earlier work focused on accelerating light transport using derivatives~\cite{Arvo94,ward1992irradiance,firstorder}.
Approximate differentiable renderers~\cite{De:2011:M3H,Loper:2014:OAD,Kato:2018:N3M,softras,nvdiffrast} have been used for many computer vision tasks,
and light transport derivatives have been used for recovering scattering coefficients~\cite{Gkioulekas:2013:IVR,Khungurn:2015:MRF}.

\vspace{0.1cm}

\noindent\textit{Background on differentiable rendering.} Much of the current interest in Monte Carlo differentiable rendering was started
by Li et al.~\shortcite{TzuMao18}, who introduced an edge sampling approach
to correctly handle discontinuities in both primary and secondary visibility.
As shown by them and subsequent work~\cite{DiffRadiativeTransfer}, the derivative of the rendering equation is made up of an interior integral which handles continuous function variation, and a boundary integral which encapsulates discontinuities.

Follow up work~\cite{Loubet,WAS,Zhang:2020:PSDR,Yan:2022:EEB,Yu:2022:EDP} focused on accurately computing the boundary integral.
Some other recent work focused on reducing memory requirements~\cite{RadiativeBackprop,Vicini2021PathReplay}, and building automatic differentiation systems and compilers~\cite{Nimier:2019:MRF,Jakob2022DrJit}.
Efforts have been made to handle different light transport phenomena~\cite{DiffRadiativeTransfer,Zhang:2021:PSD,Wu:2021:DTGR,Yi:2021:DTR}.
Much of the recent inverse rendering work has started to incorporate differentiable rendering components~\cite{Azinovic:2019:IPT,Luan:2021:USS,Nimier-David:2021:MLR,Che:2020:TLI,Deschaintre:2018:SSC}. 

~\citet{Zeltner} show that directly importance sampling a BRDF's derivative leads to a \textit{detached derivative} with only an interior term, and no boundary term.
They also show that reparameterization before differentiation leads to a different \textit{attached derivative} with not only an interior term but an additional boundary term too.
The boundary term requires careful handling for unbiased estimates and extra auxiliary rays at each shading point to estimate it too (4 to 64 extra rays as per~\citet{WAS}).
As a result, attached estimators are not suitable for the low sample budget within which we aim to operate.
Our estimators fall under the detached derivative regime, which does not require these extra auxiliary rays, which makes them suitable for low sample budget derivative estimation.

\subsection{BRDFs and Importance Sampling}

\begin{table}
\caption{
\textsc{List of Supported Material Derivatives.}
The first column lists the name of the BRDF, and the second column lists the corresponding parameter whose derivative we can importance sample.
The third column lists the type of single-signed decomposition applied (Positivization, Product Decomposition, Mixture Decomposition).
The fourth column lists the section number in the Appendix (with links) with the relevant sampling PDFs.
Please refer to the original papers for definitions of the parameters.
  }
  \setlength\tabcolsep{3.5pt}
  \begin{tabular}{ | p{5.2cm} | c | l | c |}
    \hline
    Material & Par. & SSD & PDFs \\ \hline
    Isotropic GGX~\cite{Trowbridge:1975:AIR,ggx} & $\alpha$ & Pos. & \ref{sec:app_iso_ggx} \\ \hline
    Isotropic Beckmann~\shortcite{beckmann} & $\alpha$ & Pos. & \ref{sec:app_iso_beckmann} \\ \hline
    Blinn Phong (Minnaert)~\shortcite{blinn, minnaert} & $n$ & Pos. & \ref{sec:app_blinn} \\ \hline
    Henyey-Greenstein \newline (Hanrahan-Krueger)  ~\shortcite{henyey, hanrahankrueger} & $g$ & Pos. &  \ref{sec:app_hk} \\ \hline
    Anisotropic GGX~\cite{Trowbridge:1975:AIR,ggx} & $\alpha_x, \alpha_y$ & Prod. &  \ref{sec:app_aniso_ggx} \\ \hline
    Anisotropic Beckmann (Ward) \newline ~\shortcite{beckmann, ward} & $\alpha_x, \alpha_y$ & Prod. & \ref{sec:app_aniso_beckmann} \\ \hline
    Ashikhmin-Shirley~\shortcite{ashk} & $n_u, n_v$ & Prod. & \ref{sec:app_ashk} \\ \hline
    Isotropic ABC~\cite{abc} & $B, C$ & Prod. & \ref{sec:app_abc} \\ \hline
    Isotropic Hemi-EPD~\cite{nishino} & $\kappa$ & Prod. & \ref{sec:app_epd} \\ \hline
    Burley Diffuse Reflectance~\shortcite{burley2015} & $d$ & Prod. & \ref{sec:app_bssrdf} \\ \hline
    Mixture Model (e.g., Autodesk, Disney BRDF)~\cite{autodesk, disney} & $w$ & Mix. &  \ref{sec:app_mixture} \\ \hline
    Oren-Nayar~\shortcite{orennayar} & $\sigma$ & Mix. & \ref{sec:app_on} \\ \hline
    Microcylinder~\cite{microcylinder} & $k_d$ & Mix. &  \ref{sec:app_microcylinder} \\ \hline
  \end{tabular}
  
  \label{table:brdf_list}
\end{table}

Our work supports importance sampling the derivatives of a wide variety of analytic BRDF models.
Table \ref{table:brdf_list} has a \textit{comprehensive list of the supported BRDF derivatives and their importance sampling PDFs} and the \textit{code for the sampling routines} is included in supplementary material.

\vspace{0.1cm}
\noindent\textit{Importance Sampling BRDFs.}
Importance sampling according to the BRDF~\cite{pbrt} is a fundamental variance reduction technique used in Monte Carlo forward rendering.
While essential, it was initially limited to Phong-like BRDFs~\cite{phong,Lafortune} and Ward~\shortcite{ward}.
~\citet{lawrence} introduced a non-negative matrix based factorization to efficiently fit analytic and measured BRDFs for sampling.
Walter et al.~\shortcite{ggx} introduced the GGX BRDF~\cite{Trowbridge:1975:AIR} along with its importance sampling routines.
Follow-up works have correctly accounted for the shadowing and masking terms to sample microfacet BRDFs~\cite{Heitz2018GGX, heitz2017,heitz2014, Jakob2014AnIV}.

\vspace{0.1cm}
\noindent\textit{Data-Driven BRDFs.}
Apart from analytic BRDFs, data-driven measured BRDFs~\cite{Matusik,Dupuy2018Adaptive} and Neural BRDFs~\cite{neuralbrdf, neuralbrdf2, kuznetsov2021neumip, kuznetsov2022} are another common class of BRDF models that can model a wide variety of materials.
However, both these Neural BRDFs and non-analytic measured BRDFs have a very large number of parameters, and it is unclear which parameters one might want to differentiate and importance sample with respect to.
Hence, we do not consider either of these classes of BRDFs in our work and focus instead on common analytic BRDF models.

\section{Background}
\label{sec:background}
For the sake of simplicity, we begin our discussion by focusing on the direct lighting setting, and extend it to indirect lighting in \secref{sec:gi}.
The reflected radiance $L_r$, at a shading point $\y$, in the direction $\wo$, is given by the reflection equation~\cite{cohenbook}, %
\begin{align}
    L_r(\y, \wo; \alpha) = \int f(\y, \wi, \wo; \alpha) L_i(\y, \wi) \text{d}\wi.
    \label{eq:reflection}
\end{align}
Here, $f$ is the cosine-weighted BRDF at $\y$, and $\alpha$ is a scalar BRDF parameter that controls $f$.
In practice, $\alpha$ is the vector of all BRDF parameters in a given scene.
However, for ease of exposition, we assume $\alpha$ is scalar-valued, with the results for the other parameters following similarly.
For example, $\alpha$ could be the roughness of an isotropic GGX BRDF.
Since we are dealing with only direct lighting, the incident radiance $L_i$ does not depend upon $\alpha$.
Differentiating the expression for the reflected radiance with $\alpha$, we get
\begin{align}
    \partial_\alpha L_r(\y, \wo; \alpha) = \int \partial_\alpha f(\y, \wi, \wo; \alpha) L_i(\y, \wi) \text{d}\wi.
    \label{eq:d_reflection}
\end{align}
In forward rendering, BRDF sampling aims to minimize the variance of the BRDF $f$ in the reflection equation, \eqn{\eqref{eq:reflection}}.
Similarly, our goal is to \textit{minimize the variance of the differential BRDF $\partial_\alpha f$} in differentiable rendering, which can be expressed as
\begin{align}
    I(\wo; \alpha) = \int \partial_\alpha f(\wi, \wo; \alpha) \text{d}\wi.
    \label{eq:scatter_radiance_deriv_chain_1_dbrdf}
\end{align}
We drop  the spatial coordinate $\y$, without loss of generality, for simplicity.
We deal with the incident radiance $L_i$ using light source sampling.
The estimators for $\partial_\alpha f$ and $L_i$ can be combined using Multiple Importance Sampling~\cite{Veach95MIS}.
We finally want to compute $\partial_\alpha L_r$ so the final estimator must always include multiplication by $L_i$.

\subsection{Previous Work on Variance Reduction for Differentiable Rendering}

\subsubsection{Detached \& Antithetic Sampling}
Zeltner et al.~\shortcite{Zeltner} noticed that standard BRDF sampling using a PDF $p \propto f$ for the differential BRDF $\partial_\alpha f$
leads to high variance since $\partial_\alpha f$ and $f$ can be very different functions.
They instead construct a PDF $p \propto |\partial_\alpha f|$, called the differential detached PDF, which matches $\partial_\alpha f$ in shape.
This eliminates variance from the shape of $\partial_\alpha f$, i.e., the sample weights $\partial_\alpha f / p$ are constant in magnitude.
There is, however, additional \textit{sign variance} resulting from the mismatch in the sign between the positive-valued $p$ and the real-valued integrand $\partial_\alpha f$ resulting in sample weights $\partial_\alpha f / p$ that change sign.

To deal with sign variance, Zeltner et al.~\shortcite{Zeltner} applied antithetic sampling.
While this does indeed reduce the variance of Eqn. \eqref{eq:scatter_radiance_deriv_chain_1_dbrdf}, they did not further investigate the effectiveness of antithetic sampling.
We show that Zeltner et al.'s method is a special case of another technique called positivization~\cite{OwenZhou2000}.
We show in \secref{sec:positivization} and Appendix ~\ref{sec:zeltner_anti_pos} that positivization provides a theoretical grounding of antithetic sampling: the effectiveness mainly comes from the stratification (separating the real-valued function into a positive and a negative function).
The \textit{major drawback of antithetic sampling is its inapplicability to several BRDF derivatives}, due to the lack of closed forms of root finding and integration, which we discuss in \secref{sec:positivization_anisotropic}.

\subsubsection{Antithetic Sampling of Odd Derivatives}
\citet{Antithetic} introduce another antithetic-sampling-based method to deal with the derivative of the GGX Normal Distribution Function, $D(\wh)$ with the half vector $\wh$.
They exploit the fact that the derivative $\partial_{\wh} D(\wh)$ is odd about the local shading normal, i.e
\begin{align}
    \partial_{\wh} D([\omega_{h,x}, \omega_{h,y}, \omega_{h,z}]) = -\partial_{\wh} D([-\omega_{h,x}, -\omega_{h,y}, \omega_{h,z}]).
\end{align}
Their estimator for Eqn. \eqref{eq:scatter_radiance_deriv_chain_1_dbrdf} requires two antithetic samples $\omega_{i,1}$ and $\omega_{i,2}$, and is given by
\begin{align}
    \label{eq:zhang_antithetic}
    I \approx \frac{\partial_\alpha f(\omega_{i,1}) + \partial_\alpha f(\omega_{i,2})}{p(\omega_{i,1}) + p(\omega_{i,2})}.
\end{align}
Here, and going forward, we drop $\wo, \alpha$ from the function arguments of $I(\wo, \alpha)$ and $f(\wi, \wo, \alpha)$ for brevity.
This method works well for the odd derivative with $\wh$. However, for non-odd derivatives, there are no variance reduction guarantees.
Furthermore, several BRDF derivatives are even, e.g., roughness of GGX, Beckmann, and Zhang et al.'s method increases variance in these cases.

Additionally, Eqn. \eqref{eq:zhang_antithetic} is not in the standard importance sampling form of $\partial_\alpha f / p$ due to the presence of a sum in the numerator and denominator.
Hence, it is unclear how to use it in conjunction with multiple importance sampling.

\begin{figure}[t]
  \includegraphics[width=\linewidth]{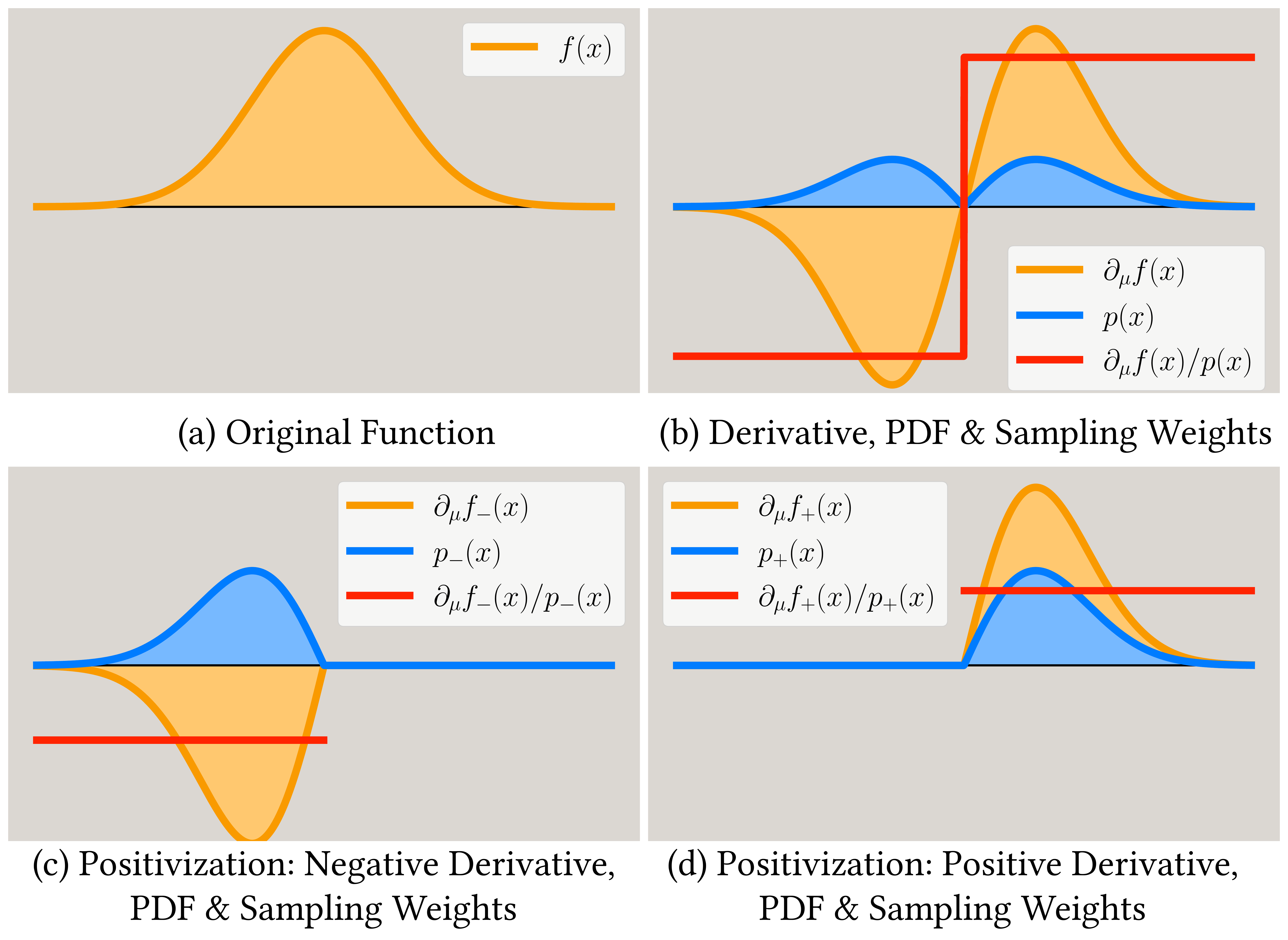}
  \vspace*{-0.7cm}
  \caption{\textsc{Sign Variance and Positivization}.
  Differentiating the positive original function (a, yellow) results in a real-valued derivative (b, yellow).
  (b) Although the derivative (b, yellow) $\partial_\mu f$ and the PDF (b, blue) $p$ match in shape, i.e., $p \propto |\partial_\mu f|$, the sample weights $\nicefrac{\partial_\mu f}{p}$ (b, red) are non-constant due to a mismatch between their signs, causing sign variance.
  Positivization~\cite{OwenZhou2000} splits the derivative into its positive and negative (c, d, yellow) parts.
  Since both parts are either purely non-negative or non-positive, they can be perfectly importance sampled by constructing PDFs $p_+$ and $p_-$ (c, d, blue).
  The resulting sampling weights (c, d, red), are constant and the corresponding estimator has zero variance. 
  }
  \label{fig:sign_variance}
  \vspace*{-0.15in}
\end{figure}

\section{single-signed Decompositions}
In this section, we describe the concept of \textit{sign variance} in real-valued integrals, and then show how our first decomposition, \textit{positivization}, can handle this source of variance for some BRDF derivatives.
Positivization requires a) analytic knowledge of roots and b) analytic integrability of the BRDF derivative, which limits its applicability.
In \secref{sec:product_rule_decomposition}, we present \textit{a novel product decomposition} that exploits the single-signed nature of the terms resulting from the product rule for derivatives, for the correct handling of sign variance.
It significantly expands the set of BRDF derivatives we can handle.
In \secref{subsec:mixture_weights}, we present \textit{a novel mixture decomposition} that exploits the fact that derivatives with mixture weights are a difference of two positive functions, to decompose them into single-signed functions, allowing us to importance sample even more BRDF derivatives.
Finally, we describe a general recipe to handle other BRDF derivatives not surveyed in this paper in \secref{sec:algorithm}.

\subsection{Sign Variance}
\label{sec:sign_related_variance}

We introduce sign variance through the following representative 1D example, showing the real-valued derivative $\partial_\mu f$ of a normal distribution $f(x; \mu, \sigma)$ with its mean $\mu$, as shown in \figref{fig:sign_variance} (a,b):
\begin{align}
    I = \int_{-\infty}^{\infty} \partial_\mu f(x; \mu, \sigma) \text{d}x
    = \int_{-\infty}^{\infty} \frac{1}{\sqrt{2\pi}\sigma^3}(x-\mu)e^{-\frac{1}{2}(\frac{x-\mu}{\sigma})^2} \text{d}x.
    \label{eq:gaussian_deriv_mu}
\end{align}
For $x < \mu$, the integrand $\partial_\mu f$ is negative, and for $x > \mu$, it is positive.
The best importance sampling strategy using a single PDF $p \propto |\partial_\mu f|$ no longer has zero variance~\cite{OwenZhou2000}.
This is due to the sign variance, i.e., the positive-valued PDF $p$ cannot match the sign of the real-valued integrand $\partial_\mu f$ over the entire domain, leading to non-constant sample weights, see \figref{fig:sign_variance} (b).

\subsection{Positivization}
\label{sec:positivization}

It is possible to construct an estimator for any real-valued integrand $\partial_\alpha f$, e.g., Eqn. \eqref{eq:gaussian_deriv_mu}, which has zero variance.
By partitioning $\partial_\alpha f$ into its positive $\partial_\alpha f_+$ and negative $\partial_\alpha f_-$ parts,
\begin{equation}
\begin{split}
    \label{eq:pos_def_split}
    \partial_\alpha f_+(x) &= \max\left(\partial_\alpha f(x), 0\right), \; \partial_\alpha f_-(x) = \min\left(\partial_\alpha f(x), 0\right) \\
    \partial_\alpha f(x) &= \partial_\alpha f_+(x) + \partial_\alpha f_-(x),
\end{split}
\end{equation}
we are left with two functions that are single-signed by definition.
They can be \textit{perfectly importance sampled if we can construct the following two PDFs}, $p_-(x) \propto \partial_\alpha f_-(x)$ and $p_+(x) \propto \partial_\alpha f_+(x)$, see \figref{fig:sign_variance}(c,d).
The resulting estimator is
\begin{equation}
\begin{split}
    I = \int \partial_\alpha f(x)\text{d}x &= \int \partial_\alpha f_+(x)\text{d}x + \int \partial_\alpha f_-(x)\text{d}x \\
    &\approx \frac{\partial_\alpha f_+(X_+)}{p_+(X_+)} + \frac{\partial_\alpha f_-(X_-)}{p_-(X_-)},
    \label{eq:pos_est}
\end{split}
\end{equation}
where $X_+ \sim p_+$ and $X_- \sim p_-$.

This technique is called positivization~\cite{OwenZhou2000}, and we apply it to importance sampling BRDF derivatives.
The zero-variance claim is only with regard to the variance arising from the differential BRDF $\partial_\alpha f$.
The derivative of the reflection equation, see Eqn. \eqref{eq:d_reflection}, is a product of the differential BRDF and the lighting, and as a result, it will still have variance from the lighting.

\begin{figure}[t]
    \includegraphics[width=\linewidth]{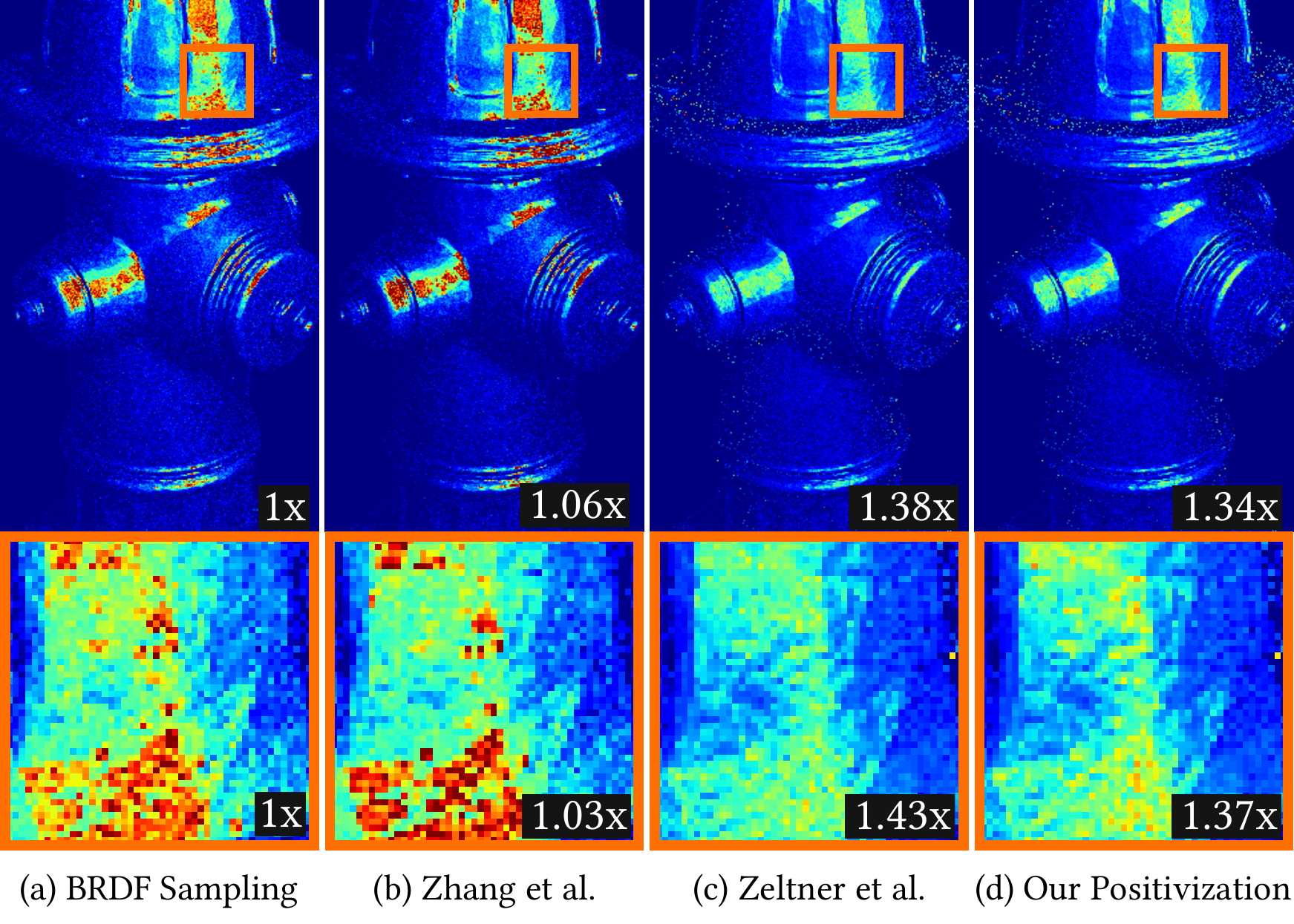}
    \caption{
Comparison between BRDF Sampling, ~\citet{Zeltner}, ~\citet{Antithetic}, and our Positivization, for the derivative of an isotropic GGX BRDF with its roughness $\alpha$.
The object in the scene is a fire hydrant lit by two area lights.
BRDF sampling is unable to correctly handle the sign or shape variance of the differential GGX BRDF and has high variance.
Zhang et al.'s method is unsuitable for even derivatives like roughness and produces a high variance estimator as well.
Zeltner et al.'s method is a special case of positivization and both of them have very similar variance reduction properties.
We have found that positivization and Zeltner et al.'s method have very similar performance across several scenes, with positivization being better in some and Zeltner et al. in others.
}
\label{fig:res_zeltner_chang}
\vspace*{-0.15in}
\end{figure}

In Appendix~\ref{sec:zeltner_anti_pos}, we show that Zeltner et al.'s antithetic sampling is a special case of positivization.
Positivization gives a theoretical grounding of Zeltner et al.'s approach.
First, through an empirical study, we have found that the majority of the variance reduction of antithetic sampling comes from the implicit splitting of $\partial_\alpha f$ into positive and negative lobes ($\partial_\alpha f_+$ and $\partial_\alpha f_-$), instead of the negative correlation between samples, see \figref{fig:res_zeltner_chang}.

\subsubsection{Positivization of Isotropic GGX}

\begin{figure}
    \centering
    \includegraphics[page=3,width=\linewidth]{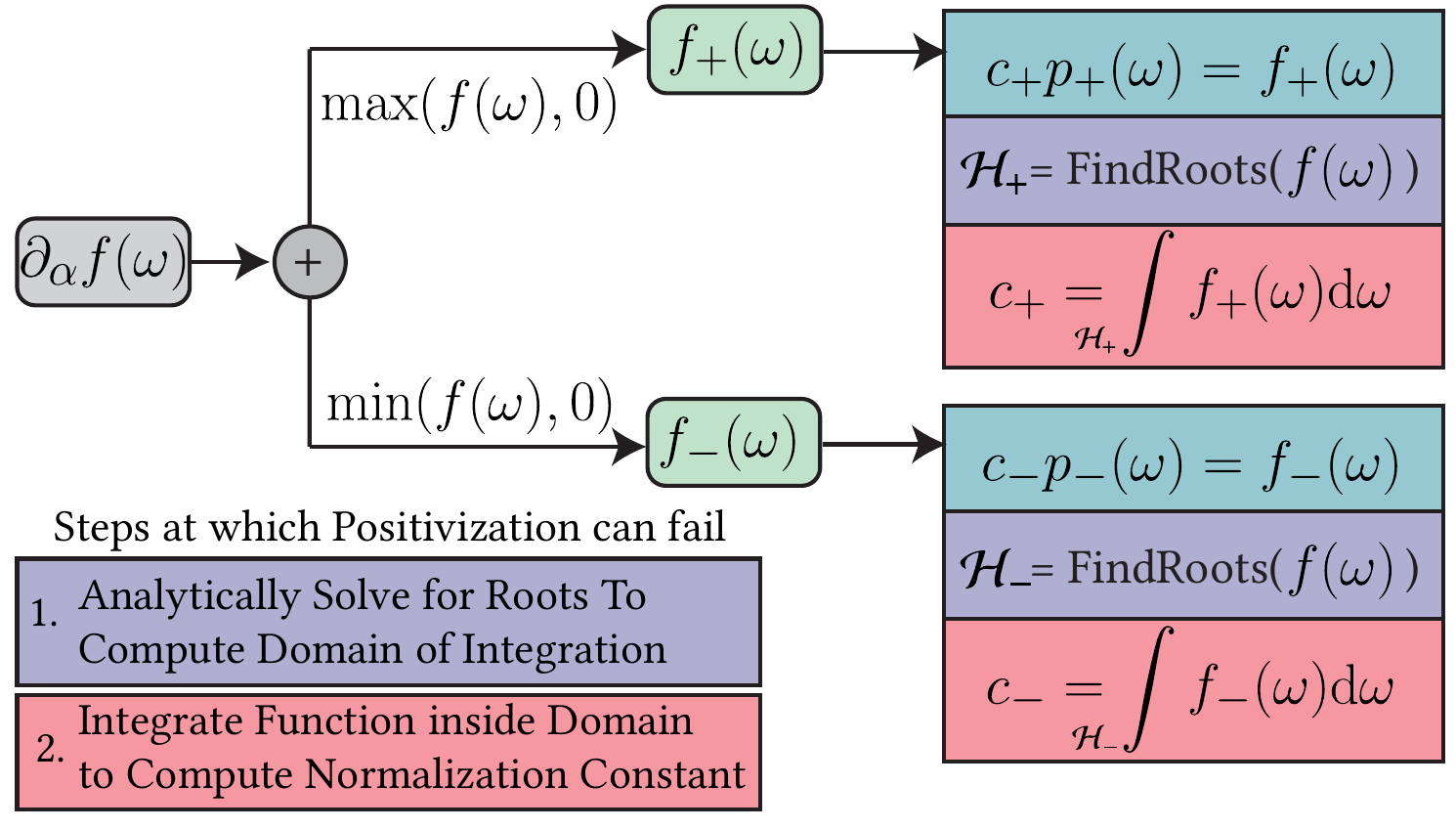}
    \caption{
    \textsc{Importance Sampling PDF construction for Positivization}.
    Positivization splits $f$ about its roots into single-signed $f_+, f_-$.
    For PDF construction, we first solve for $f$'s roots, which split the domain $\mathcal{H}$ based on $f$'s sign into $\mathcal{H_+}$ and $\mathcal{H_-}$.
    Next, we integrate inside the domain to compute the normalization constants.
    There are two potential roadblocks for positivization i) no analytic form of the roots of $f$, so $\mathcal{H_+}$ and $\mathcal{H_-}$ cannot be computed, see \figref{fig:prod_rule_bssrdf}, and ii) the positivized functions are not analytically integrable over $\mathcal{H_+}/\mathcal{H_-}$, see \figref{fig:ggx_iso_aniso_roots}.
    }
    \label{fig:pos_pipeline}
\end{figure}

Positivization, and by extension antithetic sampling, is very effective at reducing variance for BRDF derivatives when $p_+$ and $p_-$ can be constructed.
For this, $p_+$ needs to be analytically integrated over the region where $\partial_\alpha f > 0$ to obtain the necessary PDF and CDF required for sampling, see \figref{fig:pos_pipeline} for the overall pipeline to construct them.
This step faces two challenges a) the roots, which define the region where $\partial_\alpha f > 0$, don't have a closed-form expression for some BRDF derivatives, and b) $\partial_\alpha f$ isn't analytically integrable \textit{over the region} where it is positive, for others.
A similar argument follows for $p_-$ too.

Some BRDF derivatives, like isotropic microfacet GGX and Beckmann can be handled by positivization.
These microfacet models are given by the following equation,
\vspace{-0.05cm}
\begin{align}
    f(\wi, \wo) = \frac{F(\wi, \wo, \eta) G(\wi, \wo) D(\wh)}{4\cos\theta_o},
\end{align}
where $F$ is the Fresnel term, $G$ is the shadowing and masking term, and $D$ is the normal distribution function for the specific BRDF.
The unit vector $\wh$ is halfway between $\wi$ and $\wo$, and its spherical coordinates are $\theta_h, \phi_h$.

The derivative of the isotropic GGX BRDF with its roughness $\alpha$ has two components, $\partial_\alpha D(\wh)$ and $\partial_\alpha G(\wi,\wo)$. However, as noted by previous work~\cite{Zeltner, Antithetic}, the $\partial_\alpha G$ term only has a minor effect on the overall derivative.
Hence, we focus on the $\partial_\alpha D$ term, which is given by
\begin{align}
    D(\wh) &= \frac{1}{\pi\alpha^2\left( \frac{\sin^2\theta_h}{\alpha^2} + \cos^2\theta_h \right)^2} \\
    \partial_\alpha D (\wh) &= \frac{2\cos^2\theta_h\left(\tan^2\theta_h -\alpha^2\right)}{\pi\alpha^5 \left( \frac{\sin^2\theta_h}{\alpha^2} + \cos^2\theta_h \right)^3}.
\end{align}
Its roots have an analytic form and are $\tan \theta_h = \alpha$ for all $\phi_h$.
Additionally, the derivative $\partial_\alpha D$ is analytically integrable  over both the positive and negative regions, and so both conditions to apply positivization are met.
Hence, the importance sampling PDFs (and CDFs) can be obtained for this derivative.

\subsubsection{Inapplicability of Positivization to Anisotropic GGX}
\label{sec:positivization_anisotropic}

\begin{figure}
    \centering
    \includegraphics[width=\linewidth]{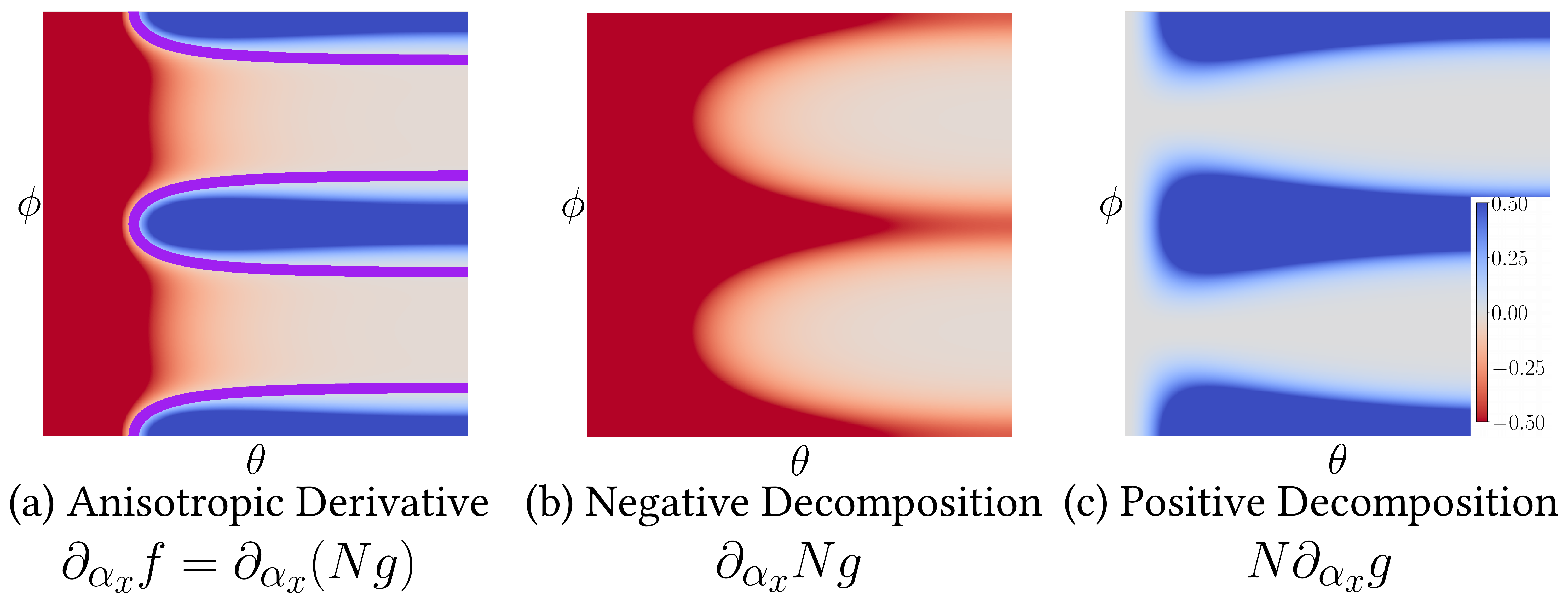}
    \caption{
    \textsc{Complicated Roots and Product Decomposition of the Anisotropic GGX BRDF derivative}.
    (a) The derivative $\partial_{\alpha_x}f$ has complicated roots shown in (a), purple curve.
    Positivization requires analytical integration over the domain defined by these roots, which is not possible.
    Our product decomposition separates the two terms resulting from the product rule for differentiation, $\partial_{\alpha_x}f = \partial_{\alpha_x}Ng + N\partial_{\alpha_x}g$, which are both single-signed as shown in (b,c).
    Product decomposition does not require integration up to the complicated roots, which enables easy PDF construction for both the positive and negative decompositions.
    }
    \label{fig:ggx_iso_aniso_roots}
\end{figure}

\begin{figure}[t]
  \includegraphics[width=\linewidth]{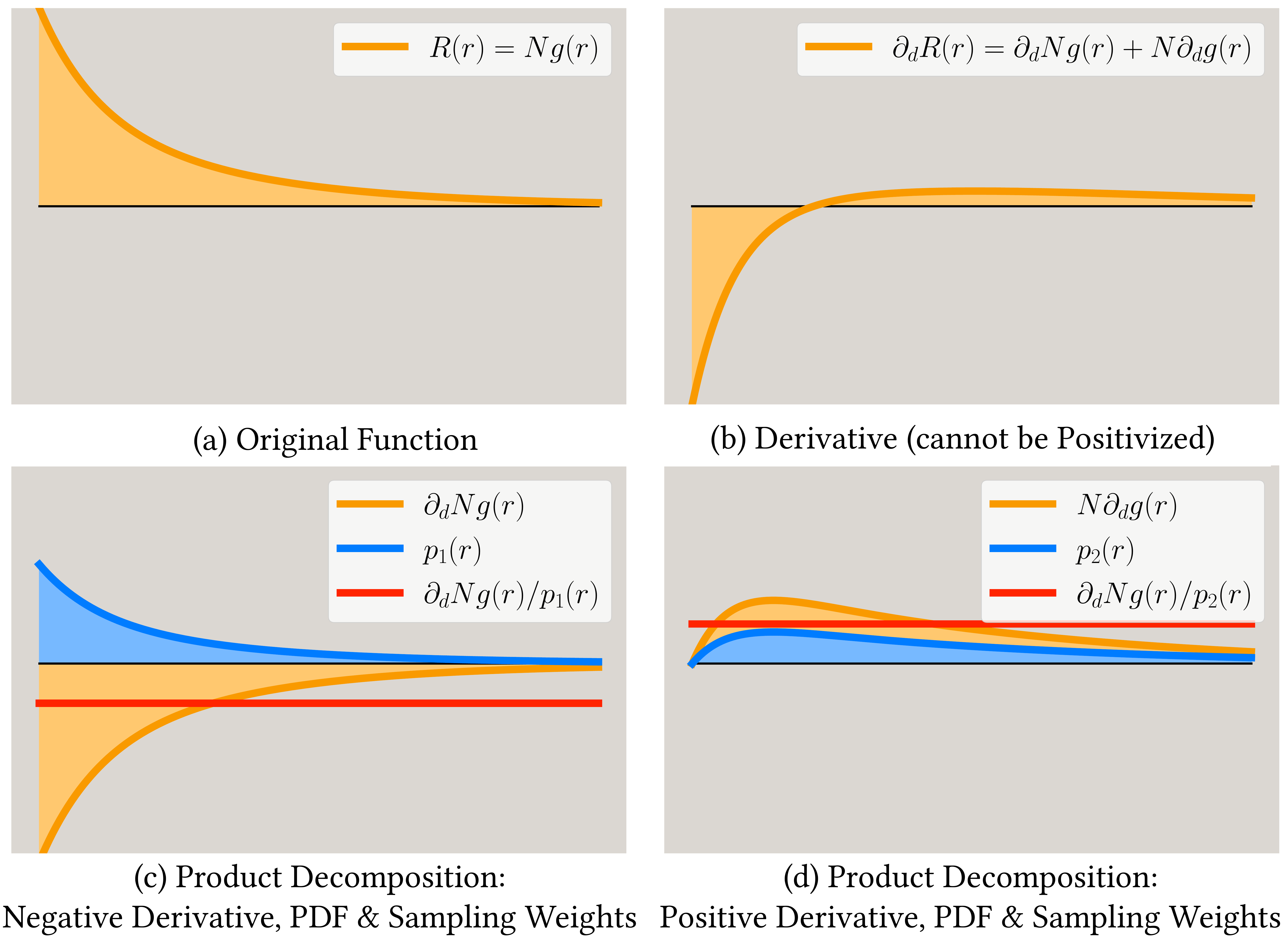}
  \vspace*{-0.7cm}
  \caption{\textsc{Product Decomposition and Inapplicability of Positivization}.
  The BSSRDF profile $R(r,d) = N(d)g(r,d)$ from ~\citet{burley2015} (a), is the product of  a shape function $g(r,d)$ and normalization term $N(d)$.
  $r$ is the spatial coordinate, and $d$ is a parameter that controls its width and height.
  Differentiating with respect to $d$ gives rise to the real-valued derivative $\partial_d R(r,d)$ (b).
  The derivative's root is given by $3e^{-2r/3d} = (3d-r)/(r-d)$, which has no analytic solution and renders positivization inapplicable to this derivative.
  However, the derivative can be written as $\partial_d R(r,d) = \partial_d N(d)g(r,d) +  N(d)\partial_d g(r,d)$ due to the product rule for derivatives.
  The first term  (c, yellow) is purely negative and the second term (d, yellow) is purely positive.
  These single-signed functions have no sign variance and can be perfectly importance sampled with the PDFs $p_1, p_2$ (c,d, blue) leading to constant sample weights (c,d, red).
  }
  \label{fig:prod_rule_bssrdf}
  \vspace*{-0.15in}
\end{figure}

For many BRDFs' derivatives, however, one of the two conditions fails,
which precludes the use of positivization (and antithetic sampling) for them.
For example, consider the derivative of $D(\wh)$ of an \textit{anisotropic} GGX BRDF with its roughness $\alpha_x$,
\begin{align}
\begin{split}
    D(\wh) &= \frac{1}{\pi\alpha_x\alpha_y\left( \frac{\sin^2\theta_h\cos^2\phi_h}{\alpha_x^2} + \frac{\sin^2\theta_h\sin^2\phi_h}{\alpha_y^2} + \cos^2\theta_h \right)^2} \\
    \partial_{\alpha_x} D(\wh) &= \frac{\cos^2\theta_h\left(3\tan^2\theta_h\cos^2\phi_h - \tan^2\theta_h\sin^2\phi_h\alpha_x^2/\alpha_y^2 - \alpha_x^2\right)}{\pi\alpha_x^4\alpha_y\left( \frac{\sin^2\theta_h\cos^2\phi_h}{\alpha_x^2} + \frac{\sin^2\theta_h\sin^2\phi_h}{\alpha_y^2} + \cos^2\theta_h \right)^3}.
    \label{eq:ggx_aniso}
\end{split}
\end{align}
Its roots are the set of $(\theta_h, \phi_h)$ such that the expression \\
$3\tan^2\theta_h\cos^2\phi_h - \tan^2\theta_h\sin^2\phi_h\alpha_x^2/\alpha_y^2 - \alpha_x^2=0$, and are shown in \figref{fig:ggx_iso_aniso_roots} (a), purple curve.
However, the derivative $\partial_{\alpha_x} D$ is \textit{not analytically integrable} over the positive and negative strata, see \figref{fig:ggx_iso_aniso_roots} (a), red and blue regions, which prevents us from applying positivization to this derivative.
Derivatives of other materials like the diffuse BSSRDF from \citet{burley2015} (\figref{fig:prod_rule_bssrdf}) and even the isotropic microfacet ABC BRDF \textit{do not have closed-form expressions for the roots}, which prevents them from being positivized.

Apart from the derivatives of the isotropic GGX, Beckmann BRDFs with their roughness, positivization can also be used for the derivative of Hanrahan-Krueger BRDF with the scattering parameter $g$ of its Henyey-Greenstein phase function.
However, like the anisotropic GGX, there are several other BRDFs, like anisotropic Beckmann, Ashikhmin-Shirley, isotropic ABC, Oren-Nayar, Burley's diffuse BSSRDF, mixture models, etc., that cannot be handled by positivization because of one of the two conditions failing.

It is common in BRDF importance sampling to numerically invert a CDF using binary search or Newton iterations, and we will do this with some of our product and mixture decomposition CDFs.
However, for positivization, taking a purely numerical approach is not practical.
Numerically approximating non-analytic roots and non-analytically integrable PDFs requires storing a high dimensional representation (for e.g. 6D $\wi, \wo, \alpha_x, \alpha_y$ for anisotropic GGX).
Storing such a high dimensional histogram (piecewise approximation) can be infeasible.

Positivization is a specific single-signed decomposition that decomposes the real-valued function into a positive and a negative function \textit{with non-overlapping supports}.
As a result, it requires root-finding and analytic integration over complicated domains defined by these roots.
In the following two sections, we discuss two novel decompositions for which the positive and negative functions are defined over simple domains of integration like a plane or hemisphere, with overlapping support.
As a result, they \textit{do not require root finding, or integration over complicated domains}, and as a consequence can handle a broader class of derivatives.

\section{Product Decomposition}

\label{sec:product_rule_decomposition}

\begin{figure}
    \centering
    \includegraphics[page=5,width=\linewidth]{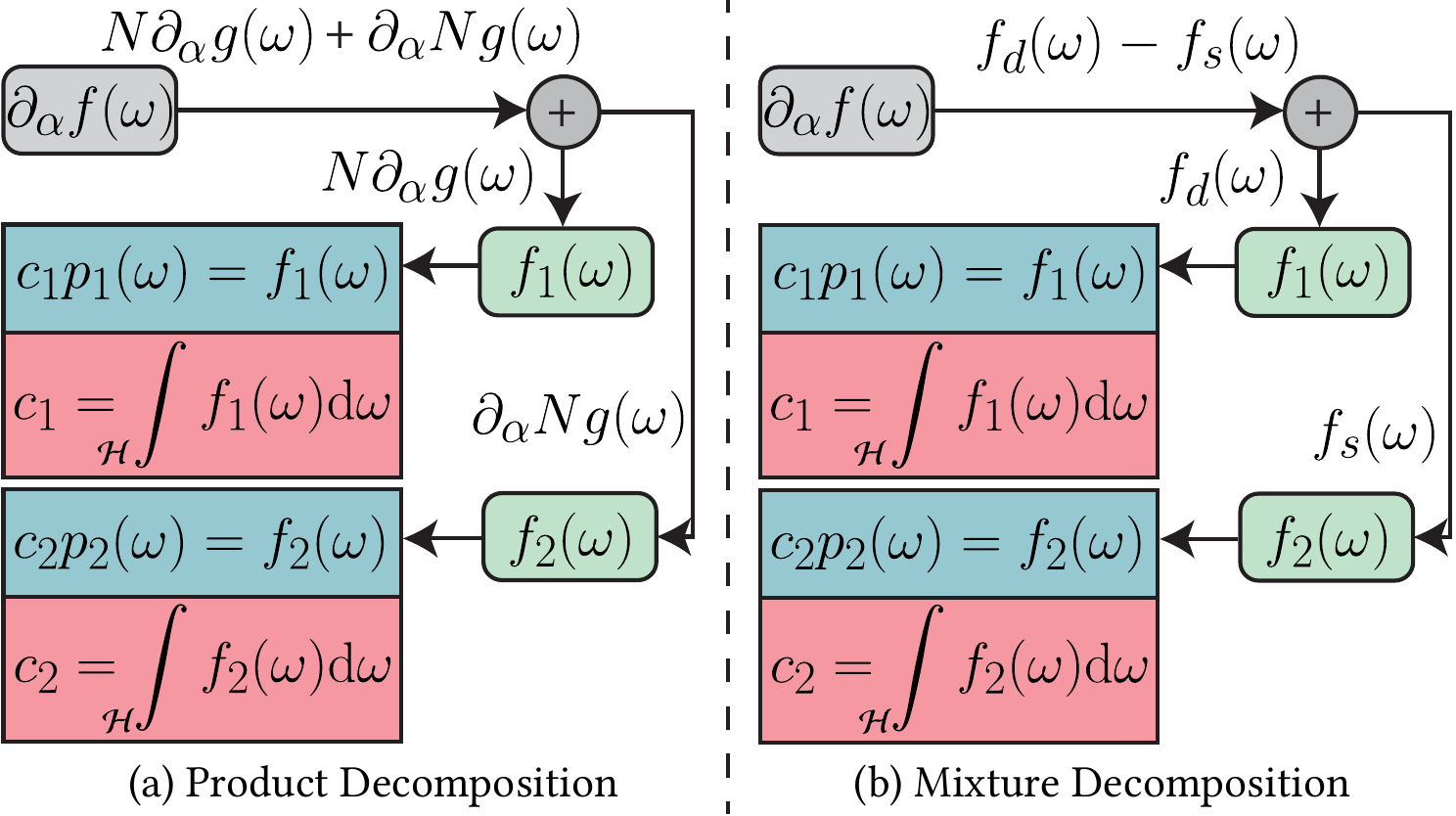}
    \caption{
    \textsc{Importance Sampling PDF construction for Product and Mixture Decomposition}.
    Differentiating the positive-valued functions $f(\omega, \alpha) = N(\alpha)g(\omega, \alpha)$ gives us $\partial_\alpha f(\omega, \alpha) = \partial_\alpha Ng(\omega) +  N\partial_\alpha g(\omega)$ for product decomposition, and differentiating $f(\omega, \alpha) = \alpha f_d(\omega) + (1 - \alpha) f_s(\omega)$ gives us  $\partial_\alpha f(\omega, \alpha) =  f_d(\omega) -  f_s(\omega)$ for mixture decomposition.
    For both decompositions, the two terms $f_1(w)$ and $f_2(\omega)$ are single-signed.
    Neither of the decompositions requires complicated root finding and integration up to the roots, unlike positivization, see \figref{fig:pos_pipeline}.
    Instead, the integration domain is simply the entire hemisphere $\mathcal{H}$.
    This makes PDF construction for these decompositions possible for several BRDF derivatives where positivization was not applicable.
    }
    \label{fig:prod_mix_pipeline}
    
\end{figure}

Our first new decomposition is product decomposition. It can handle the derivatives of anisotropic microfacet BRDFs, diffuse BSSRDFs, and the isotropic ABC BRDF that positivization could not handle.
The key idea we exploit is that after differentiating any of these materials, they split up into two terms following the product rule.
Both of these are single-signed, have no sign variance, and are analytically integrable over their simple domains of integration (hemisphere or plane).

Several BRDFs (or normal distribution functions) are of the form
\begin{align}
    f(\wh, \alpha) = N(\alpha) g(\wh, \alpha),
\end{align}
where $g(\wh, \alpha)$ is a non-negative shape function, which determines the overall shape of the BRDF over all $\wh$, at the parameter value $\alpha$.
$N(\alpha)$ is a directionally constant (independent of $\wh$) normalization term that ensures $f$ integrates to $1$.
Differentiating $f$ with $\alpha$, we get,
\begin{align}
    \partial_\alpha f(\wh, \alpha) = \partial_\alpha N(\alpha) g(\wh, \alpha) + N(\alpha) \partial_\alpha g(\wh, \alpha).
\end{align}
\textit{Because $N$ and $\partial_\alpha N$ are directionally constant, the variance in the two terms above comes from $g$ and $\partial_\alpha g$ respectively}.
The first term above is single-signed because $g \geq 0$.
The second term with $\partial_\alpha g$ can potentially be real-valued.
However, we have found it to be single-signed for several common BRDFs.
For example, for the anisotropic GGX normal distribution function $D(\wh)$, we have,
\begin{equation}
\begin{split}
    D(\wh,\alpha_x,\alpha_y) &= N(\alpha_x,\alpha_y)g(\wh, \alpha_x,\alpha_y)\\
    N(\alpha_x, \alpha_y) &= (\pi \alpha_x \alpha_y)^{-1} \\
    g(\wh, \alpha_x,\alpha_y) &=
    \left( \frac{\sin^2\theta_h\cos^2\phi_h}{\alpha_x^2}
    + \frac{\sin^2\theta_h\sin^2\phi_h}{\alpha_y^2} + \cos^2\theta_h \right)^{-2} \\
    \partial_\alpha g(\wh, \alpha_x,\alpha_y) &= 4 g(\wh, \alpha_x,\alpha_y)^{3/2} \sin^2\theta_h\cos^2\phi_h \alpha_x^{-3},
\end{split}
\end{equation}
where $\partial_\alpha g$ is single-signed, see \figref{fig:ggx_iso_aniso_roots}.
Additionally, $\partial_\alpha g$ is also analytically integrable over its hemispherical domain.

Let us provide some geometric intuition for why the shape derivative $\partial_\alpha g$ is often single-signed. For our BRDFs, the parameter $\alpha$ often controls the variance of the distribution, e.g., $\alpha_x, \alpha_y$ for GGX, Beckmann, $n_u, n_v$ for Ashikhmin-Shirley.
For all of these, the variance $\alpha$ stretches $g$ horizontally, and increases (or decreases) its value at all locations, making its derivative single-signed.
On the other hand, $\alpha$ stretches $N(\alpha)$ vertically to negate the increase (or decrease) in area due to $g$, and ensure it integrates to $1$.

We construct importance sampling PDFs for the two single-signed terms separately, with PDFs $p_1 \propto  g$ and $p_2 \propto  \partial_{\alpha}g$,
\begin{equation}
\begin{split}
    I &= \int \partial_{\alpha}f(\wh)\text{d}\wh \\
    &= \int \partial_{\alpha}Ng(\wh)\text{d}\wh + 
    \int N\partial_{\alpha} g(\wh) \text{d}\wh \\
    &\approx \frac{\partial_{\alpha}Ng(\whone)}{p_1(\whone)} + \frac{N\partial_{\alpha} g(\whtwo)}{p_2(\whtwo)}.
    \label{eq:prod_est}
\end{split}
\end{equation}
\figref{fig:prod_mix_pipeline} (a) describes the pipeline to generate importance sampling PDFs for product decomposition.
Product decomposition can handle the derivatives of anisotropic GGX, Beckmann, Ashikhmin-Shirley which are not analytically integrable over the positivized domains, and Burley's diffuse BSSRDF and the isotropic ABC BRDF, which have no closed-form solution for the roots. However, they all have single-signed $\partial_\alpha g$ which is analytically integrable.

Note that the product rule in and of itself does not guarantee a single-signed decomposition.
For example, the product of the microfacet distribution ($D$) and geometric terms ($G$) does not lead to a single-signed decomposition for the derivative with $\alpha_x$ (or $\alpha_y$).
This is because both $\partial_{\alpha_x} D$ and the $\partial_{\alpha_x} G$ terms are real-valued. 
The decomposition $D = N g$ is one of the many product decompositions, but the only one we found to preserve the single-signed property.

\section{Mixture Decomposition}
\label{subsec:mixture_weights}
Our second new decomposition further expands the set of BRDF derivatives we can handle.
Consider, for example, a BRDF made up of a diffuse $f_d$ and specular $f_s$ lobe with scalar mixture weights $k_d$ and $1-k_d$ respectively.
\begin{equation}
\begin{split}
    f(\wi, \wo) &= k_d f_d(\wi, \wo) +
    (1 - k_d) f_s(\wi, \wo) \\
    \partial_{k_d} f(\wi, \wo) &= f_d(\wi, \wo) - f_s(\wi, \wo).
\end{split}
\label{eq:mix_deriv}
\end{equation}
The derivative with the mixture weight $k_d$ is positive when the diffuse lobe contribution is higher than the specular lobe and negative otherwise.
In general, this derivative is very hard to positivize, because $f_d$ and $f_s$ can be arbitrary BRDFs, and so the roots of $f_d - f_s$ are unlikely to have a simple analytic form.

However, we can once again decompose this derivative into single-signed functions with overlapping support; we refer to this as the mixture decomposition.
Since $f_d$ and $f_s$ are non-negative valued BRDFs, they are single-signed, and can be importance sampled separately with appropriate PDFs $p_d$ and $p_s$.

\begin{equation}
\begin{split}
    I &= \int \partial_{k_d} f(\wi) \text{d}\wi = \int \partial_{k_d} f_d(\wi) \text{d}\wi - \int \partial_{k_d} f_s(\wi) \text{d}\wi \\
    &\approx \frac{f_d(\wid)}{p_d(\wid)}
    - \frac{f_s(\wis)}{p_s(\wis)}.
\end{split}
\label{eq:mix_est}
\end{equation}
Mixture weights show up in all \emph{Uber} BRDFs, like the Autodesk Standard Surface, Disney BRDF, etc., and our mixture decomposition can be applied to all of them.

Mixture decomposition is also applicable to the derivative of BRDFs that aren't explicitly mixture models, but internally are made up of different lobes, with parametric weights.
For example, the Oren-Nayar BRDF, which is a linear combination of two terms.
Here, the positive weights $A(\sigma), B(\sigma)$ depend upon the roughness $\sigma$ of the BRDF.
\begin{equation}
\begin{split}
    f(\wo, \wi) &= A(\sigma)\frac{\rho}{\pi}\cos\theta_i \\
    &+ B(\sigma) \frac{\rho}{\pi}\max\left(0, \cos(\phi_i-\phi_o)\right)\sin\alpha\tan\beta\cos\theta_i,
\end{split}
\label{eq:oren_nayar}
\end{equation}
where $\alpha=\max\left(\theta_i,\theta_o\right)$, $\beta=\min\left(\theta_i,\theta_o\right)$.
Once again, since both terms of the BRDF above are positive, the real-valued derivative with $\sigma$ is simply the sum of a positive and a negative term,
\begin{align}
\begin{split}
    \partial_\sigma f(\wo, \wi) &= \partial_\sigma A(\sigma)\frac{\rho}{\pi}\cos\theta_i \\
    &+ \partial_\sigma B(\sigma) \frac{\rho}{\pi}\max\left(0, \cos(\phi_i-\phi_o)\right)\sin\alpha\tan\beta\cos\theta_i, \label{eq:on_deriv}
\end{split}
\end{align}
with the sign of the term decided by the sign of $\partial_\sigma A$ and $\partial_\sigma B$.
Importance sampling the first term is simply cosine-hemispherical sampling, and we provide an importance sampling PDF for the second term in Appendix~\ref{sec:app_on}.
Besides Oren-Nayar, the microcylinder BRDF~\cite{microcylinder} is also a mixture model with weights $k_d, 1 - k_d$, where $k_d$ is the isotropic scattering coefficient, and can be handled by mixture decomposition as well.

\section{Recipe for importance sampling BRDF derivatives}

\begin{figure}
    \centering
    \includegraphics[width=\linewidth]{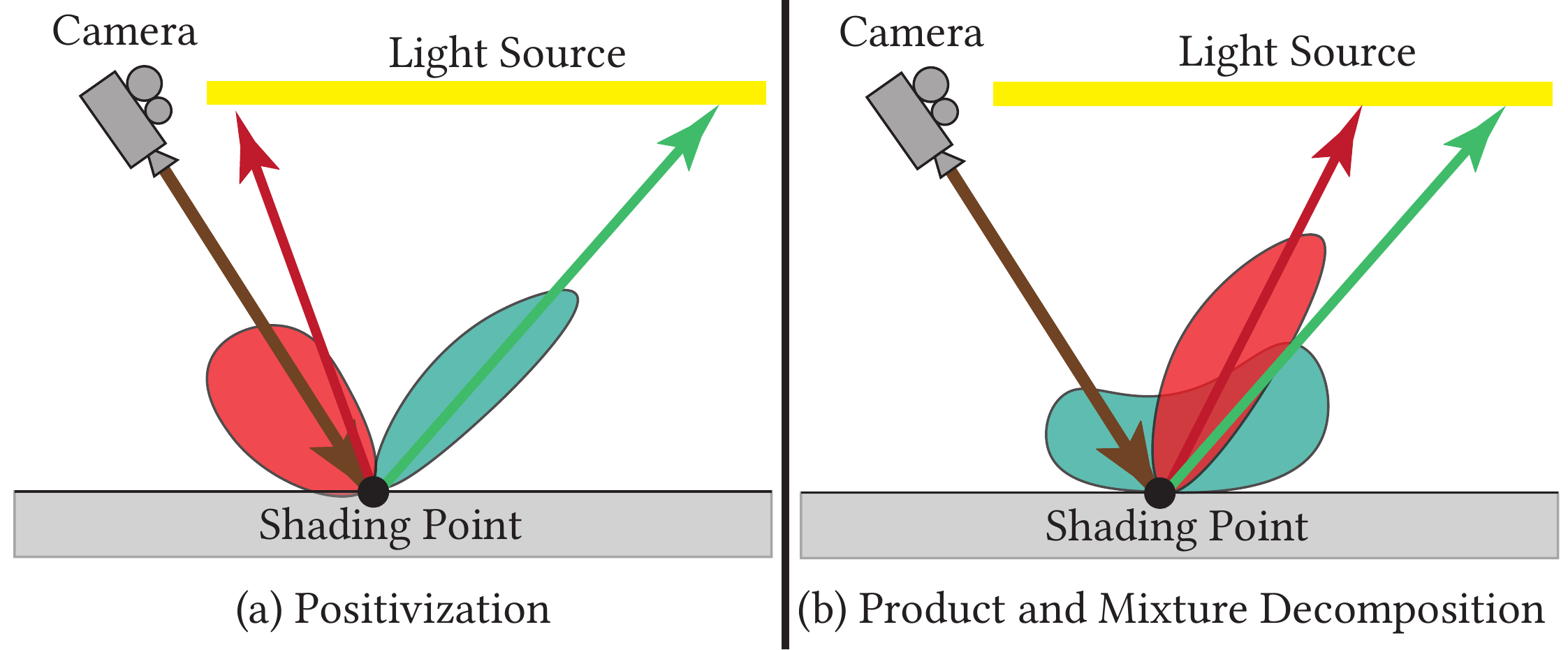}
    \caption{
    Positivization, Product and Mixture Decomposition for Direct Illumination.
    All three techniques send out two shadow rays corresponding to two different sampling techniques at each shading point, shown by red and green arrows.
    For positivization, the PDFs for these sampling techniques have non-overlapping support, shown by the red and green lobes.
    For mixture and product decomposition, however, the corresponding PDFs may have overlapping support.
    }
    \label{fig:pos_prod_mix_di_cartoon}
\end{figure}

\label{sec:algorithm}
We now present a \emph{recipe} to importance sample BRDF derivatives based on the key ideas introduced in the previous sections.

\vspace{0.2cm}
\noindent\textbf{Step 1, Positivization.} Given a real-valued BRDF derivative $\partial_\alpha f$, check if it can be positivized.
For positivization to be applicable, $\partial_\alpha f$ should have analytic roots.
Compute the normalization constants for the solid angle PDFs $p_+(\wi) \propto \max\left(\partial_\alpha f(\wi), 0\right),\; p_-(\wi) \propto \min\left(\partial_\alpha f(\wi), 0\right)$, and their marginal and conditional counterparts\\ $p_+(\phi_i), p_-(\phi_i), p_+(\theta_i|\phi_i), p_-(\theta_i|\phi_i)$, if they are analytically integrable.
See \figref{fig:pos_pipeline} for the PDF generation, and Eqn. \eqref{eq:pos_est} for the estimator.

\vspace{0.2cm}
\noindent\textbf{Step 2, Try Product or Mixture Decomposition.} If positivization is inapplicable for either reason (no analytic roots or lack of analytic integrability), try to apply either product or mixture decomposition.

\vspace{0.2cm}
\noindent\textbf{Step 2.1, Product Decomposition.} If the original BRDF is of the form $N(\alpha)g(\wi,\alpha)$, where $\alpha$ appears in a directionally invariant (independent of $\wi$) normalization term $N(\alpha)$ and an unnormalized shape function $g(\wi, \alpha)$, product decomposition may be applicable.
First check if $\partial_\alpha g$ is single-signed, i.e., it has a constant sign for all $\wi$, and is analytically integrable.
If these conditions hold, product decomposition is applicable. Construct a PDF $p_2(\wi) \propto \partial_\alpha g$ and compute the normalization terms for it and its conditional and marginal counterparts.
The other PDF $p_1(\wi) \propto g$ is simply the BRDF sampling PDF.
See \figref{fig:prod_mix_pipeline} for the PDF generation, and Eqn. \eqref{eq:prod_est}, for the estimator.

\vspace{0.2cm}
\noindent\textbf{Step 2.2, Mixture Decomposition.} If instead the parameter $\alpha$ appears in the form of linear combination weights either explicitly as a mixture model between two BRDFs, or implicitly as a mixture between two lobes that form a single BRDF, mixture decomposition is likely applicable here.
In this case, simply use the PDFs and sampling strategies most suitable for the two mixture lobes if they are available (e.g., visible normal distribution function sampling for a GGX lobe), or construct PDFs $p_1(\wi) \propto f_1(\wi),\; p_2(\wi) \propto f_2(\wi)$ for the two lobes, where $f_1, f_2$ are the two lobes. 
See \figref{fig:prod_mix_pipeline} for the PDF generation, and Eqn.~\eqref{eq:mix_est} for the estimator.

\figref{fig:pos_prod_mix_di_cartoon} depicts the estimators for all three of our decompositions for direct illumination.
They all require two shadow rays at the shading point, corresponding to the positive and negative lobes of the corresponding decomposition.

Although we have not found examples that require it, our three decompositions can also be interleaved with one another for complicated BRDF derivatives.
For example, it is possible that for some BRDF derivatives, the derivative of the shape function from the product rule $\partial_\alpha g$ could be real-valued.
It could then further be positivized to eliminate sign variance.

\vspace{0.1cm}
\noindent\textit{Forward Rendering Sampling Technique Reuse.}
Both product and mixture decomposition reuse BRDF sampling developed for forward rendering as one (or both) of the techniques for differential BRDF sampling.
For product decomposition, this corresponds to $p_1 \propto g$.
For mixture decomposition, perfect importance sampling can be achieved by \textit{only} employing two standard BRDF sampling techniques from forward rendering in some cases.
BRDF sampling when used directly to estimate for $\partial_\alpha f$ suffers from sign and shape variance, however, when paired with the right decomposition, it can correctly handle the shape variance of one of the terms.

\vspace{0.1cm}
\noindent\textit{Multiple Importance Sampling.}
For the product and mixture decompositions, the positive and negative decomposition PDFs can have overlapping support (for positivization they are necessarily non-overlapping).
As a result, the samples generated for one decomposition can be shared with the other using Multiple Importance Sampling. 
Also, all three of our decompositions reduce the variance from the differential BRDF $\partial_\alpha f$, and can be used in conjunction with light source sampling via Multiple Importance Sampling to reduce the lighting, $L_i$'s variance.

\section{Results}
\label{sec:results}

We organize our results into two subsections. First, we demonstrate that our decompositions do reduce variance in practice for a number of BRDF derivatives under a wide variety of lighting conditions in \secref{sec:res_var_red}.
Next, we demonstrate that lower variance in gradients indeed does enable better spatially-varying texture recovery in an inverse rendering setting, in \secref{sec:res_inv_rend}.

\vspace{0.1cm}
\noindent\textit{Implementation Details.}
We implemented all the different decompositions and BRDFs on our own CPU-based
differentiable renderer, using the Embree ~\cite{embree} library for ray tracing.
At each shading point, all three of our decompositions require two shadow rays, see \figref{fig:pos_prod_mix_di_cartoon}.
To have a fair comparison with BRDF sampling, we shoot out two shadow rays at each shading point for it too, which ensures an equal-ray comparison with our method.
All our error comparison images are computed by averaging the squared error of the gradient images, which were each generated at 9 samples per pixel over 50 runs.
We use numerical CDF inversion for the sampling of $\partial_\alpha g$ in product decomposition and the $\partial_\sigma B (\sigma)$ term in Oren-Nayar.
\subsection{Variance Reduction}

\label{sec:res_var_red}
\subsubsection{Positivization}

First, we compare positivization with BRDF sampling for the derivative of two BRDFs in \figref{fig:teaser}.
The scene is lit by two area lights.
The isotropic GGX teapot (with $\alpha=0.02$) is differentiated with its roughness $\alpha$, and the Hanrahan-Krueger (with $g=-0.9$) lion is differentiated with its Henyey-Greenstein parameter for anisotropy $g$.
The Henyey-Greenstein phase function at $g=-0.9$ is highly back-scattering and is very badly importance sampled by regular BRDF sampling, which cannot correctly account for the highly peaked and signed nature of the derivative.
Since positivization is correctly able to handle both sign and shape related variance, we see significant variance reduction of $1.96\times$ and $58.57\times$ for the teapot and lion respectively.

\subsubsection{Product Decomposition}
\begin{figure}[t]
    \includegraphics[width=\linewidth]{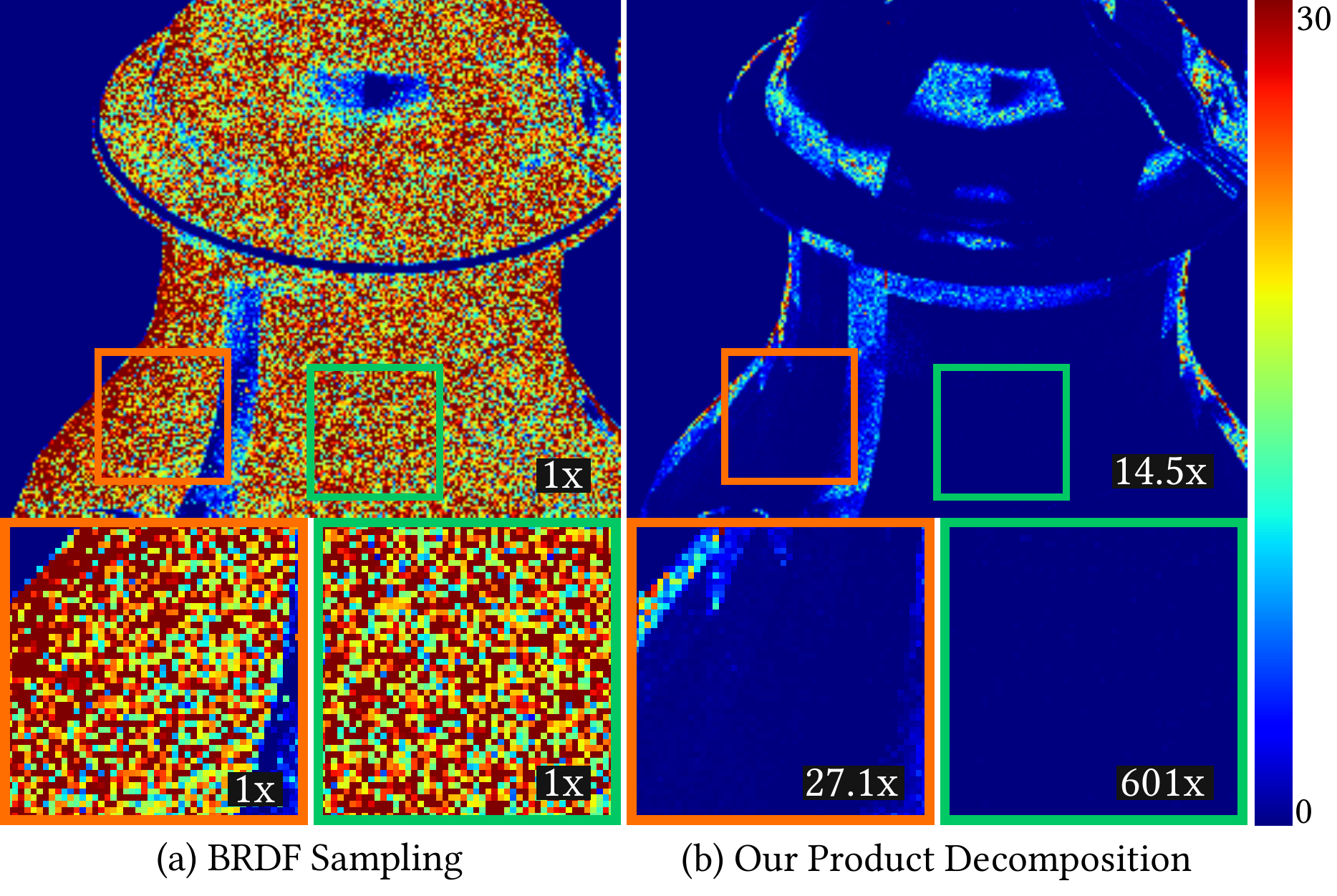}
    \vspace*{-0.8cm}
    \caption{\textsc{Product decomposition v.s. BRDF sampling under constant illumination}. 
    We show the estimated variance of derivatives. Numbers indicate the relative improvement, \textbf{higher is better}. The scene contains an anisotropic Beckmann BRDF under constant environment illumination.
    Under constant illumination, the BRDF derivative is the main source of variance. Our product decomposition correctly handles both the sign and shape variance, because of which we see an overall $14.5\times$ reduction in variance compared to BRDF sampling.
    }
    \label{fig:brdf_prod_beckmann_const}
    \vspace*{-0.15in}
\end{figure}
Next, we compare product decomposition with BRDF sampling for the derivative of an anisotropic Beckmann BRDF with its roughness $\alpha_x$, lit under constant environment illumination in \figref{fig:brdf_prod_beckmann_const}.
Positivization (and by extension Zeltner et al.) cannot handle this derivative, see \secref{sec:positivization_anisotropic}, and Zhang et al.'s method fails for even derivatives like this one, see \figref{fig:res_zeltner_chang}.
Constant illumination eliminates variance from lighting and only keeps variance from the BRDF derivative and visibility.
Since product decomposition can correctly handle both the sign and shape variance of the BRDF derivative, it has an overall $14.5\times$ reduction in variance, whereas BRDF sampling fails because it cannot handle either source of variance.
In most regions (\figref{fig:brdf_prod_beckmann_const} see right inset), the derivative of the normal distribution function $\partial_\alpha D$ is the major source of BRDF derivative variance; we eliminate it and see a big improvement of $601\times$.
However, in the grazing angle regions (\figref{fig:brdf_prod_beckmann_const} see left inset), the derivative of the shadowing function $\partial_\alpha G$ dominates.
Here, our improvement is still significant ($27.1\times$), but relatively less pronounced, since our sampling strategy minimizes $\partial_\alpha D$'s variance.

\begin{figure}[t]
    \includegraphics[width=\linewidth]{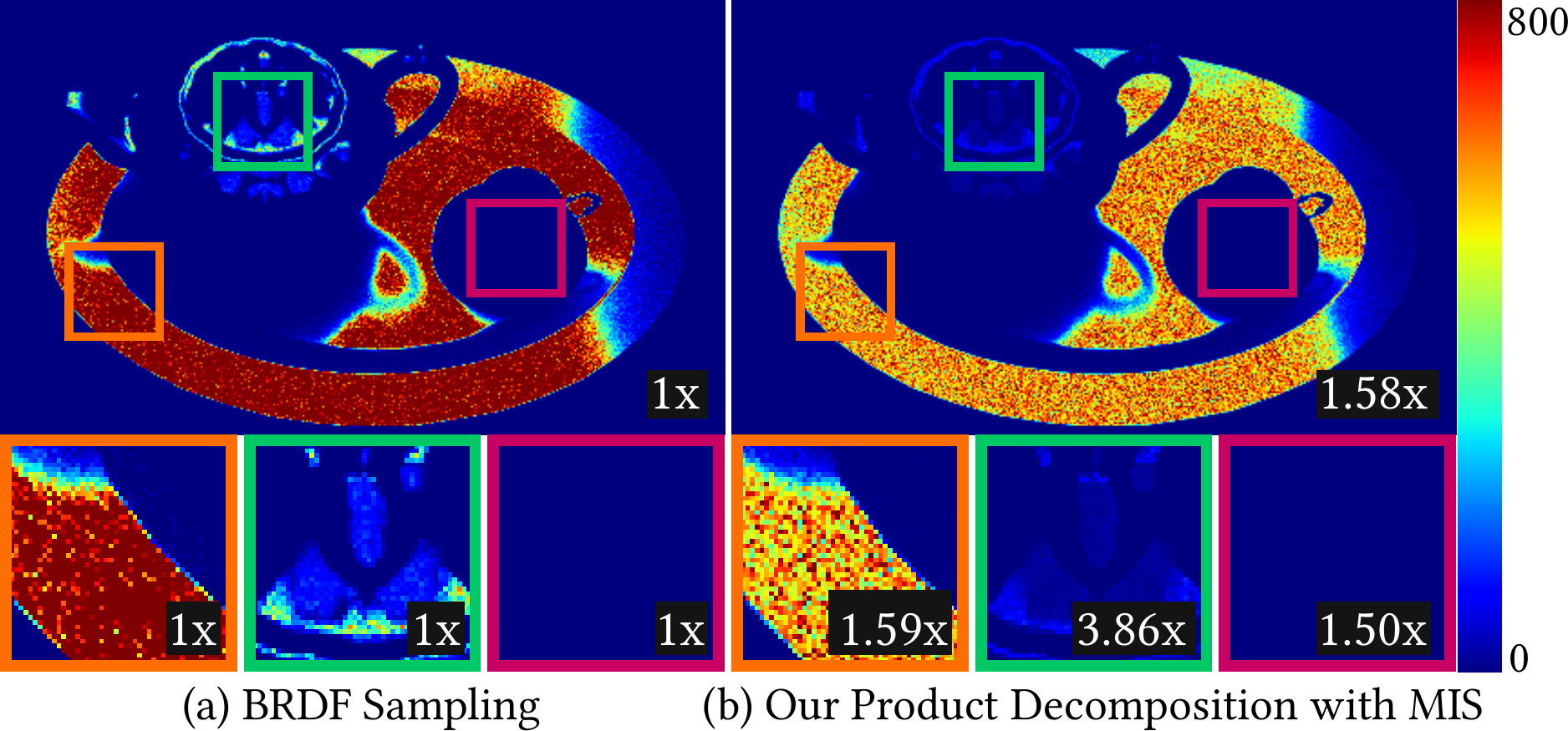}
    \caption{\textsc{Product decomposition v.s. BRDF sampling under sharper illumination}. We show the estimated derivative variance improvement for three anisotropic microfacet BRDFs, \textbf{higher is better}, for our product decomposition with MIS compared with BRDF sampling.
    The BRDFs are GGX for the tray, Beckmann for the pot, and Ashikhmin-Shirley for the cup, and are lit by two area lights.
    The visibility (from the object inter-occlusions) and the lighting introduces additional variance from these terms.
    Nontheless, our product decomposition which shares samples across the positive and negative lobes using MIS, is once again better able to handle the sign and shape variance than BRDF sampling, leading to an improvement of $1.58\times$.
    }
    \label{fig:res_prod_aniso_all}
    \vspace*{-0.15in}
  \end{figure}

Now, we change the lighting to a realistic setup with two area lights and three anisotropic BRDFs, in \figref{fig:res_prod_aniso_all}.
The three anisotropic BRDFs are GGX for the tray, Beckmann for the pot, and Ashikhmin-Shirley for the cup, and we compute the derivatives with $\alpha_x$ for GGX and Beckmann, and $n_u$ for Ashikhmin-Shirley.
The Ashikhmin-Shirley gradient is scaled up by $10^3$ since it has a lower magnitude.
Apart from BRDF derivative variance, this scene has two other major sources of variance, lighting, and visibility.
When the variance is significant from other sources too, we have found that sharing samples between the positive and negative decomposition is beneficial, see \secref{sec:algorithm}, Multiple Importance Sampling (MIS).
Our product decomposition with MIS better handles the shape and sign variance than BRDF sampling, and can outperform it for all three BRDF derivatives (see three insets in \figref{fig:res_prod_aniso_all}), and has an overall variance reduction of $1.58\times$.

We show two more examples of product decomposition in \figref{fig:teaser}, for anisotropic GGX and Beckmann BRDF derivatives, which achieve variance reduction of $1.56\times$ and $3.61\times$ respectively.
The insets in the top row of \figref{fig:teaser} show the regions where our decomposition has lower variance than BRDF sampling in blue. Product decomposition outperforms BRDF sampling in almost all regions.

\subsubsection{Mixture Decomposition}
Finally, we compare BRDF sampling with Mixture Decomposition to estimate the derivative of a mixture model with its mixture weight for the fish-shaped pot in \figref{fig:teaser}.
The mixture model is a linear combination of a lambertian diffuse lobe, and a GGX specular lobe and the lighting is two area lights.
Mixture decomposition can reduce the variance by 4.72x, because it correctly handles shape and sign variance, unlike BRDF sampling.

\figref{fig:teaser} also shows an example of an Oren-Nayar pot, and its derivative with the roughness $\sigma$.
BRDF sampling here is simply cosine hemispherical sampling, and works quite well in the central regions of the pot, because the cosine lobe is dominant in the non-grazing angle regions, see Eqn. \eqref{eq:oren_nayar}.
However, in the grazing angle regions towards the edges of the pot where the correction term is more dominant, BRDF sampling breaks down and has high variance.
On the other hand, our mixture decomposition with MIS correctly accounts for the derivative of both terms with regard to their sign and shape variance, and can achieve low variance in \textit{all} regions of the pot, and leads to a $3.91\times$ reduction in variance.

\subsection{Inverse Rendering}
\label{sec:res_inv_rend}

\begin{figure}[t]
    \includegraphics[width=\linewidth]{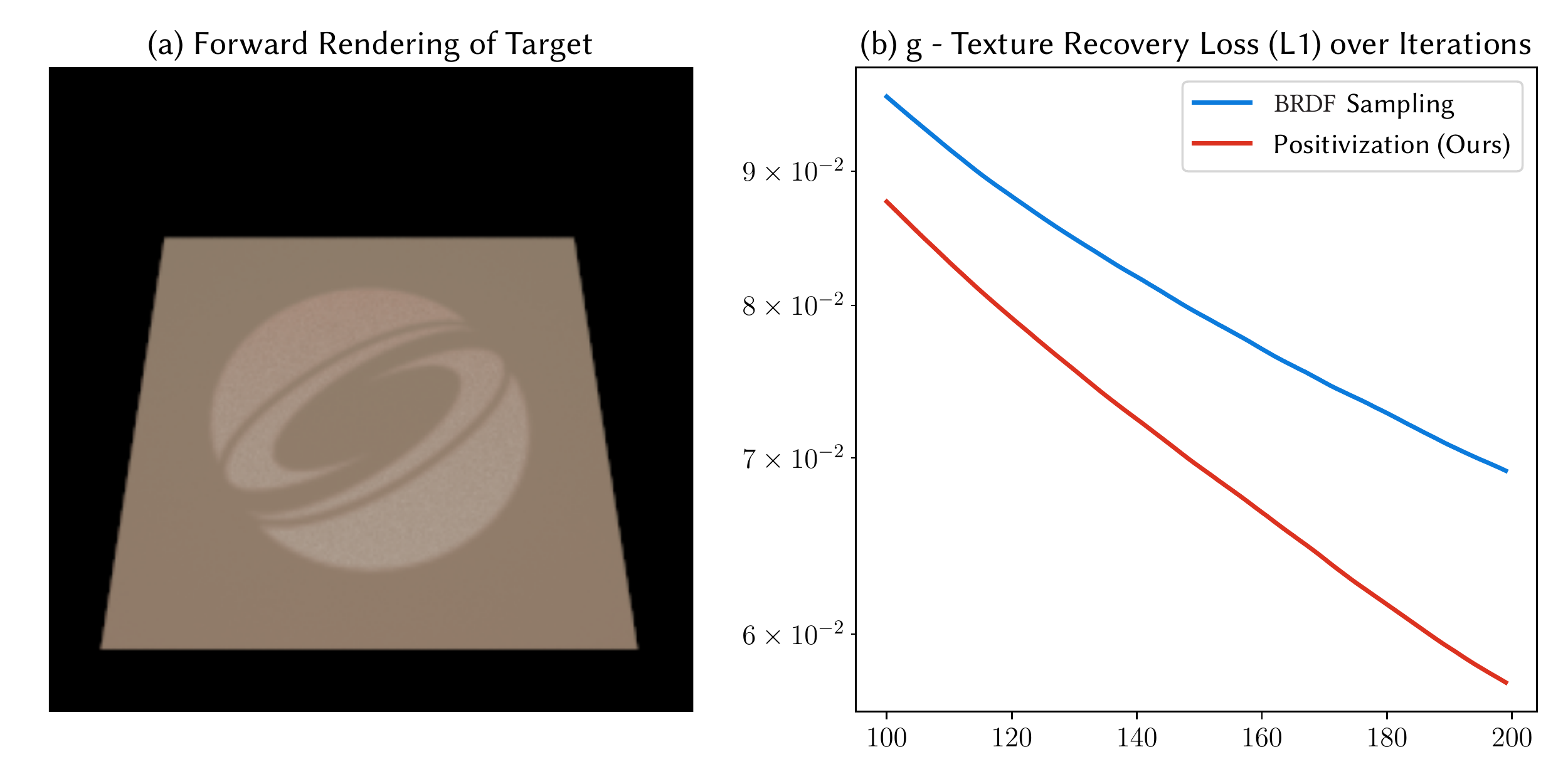}
    \includegraphics[width=\linewidth]{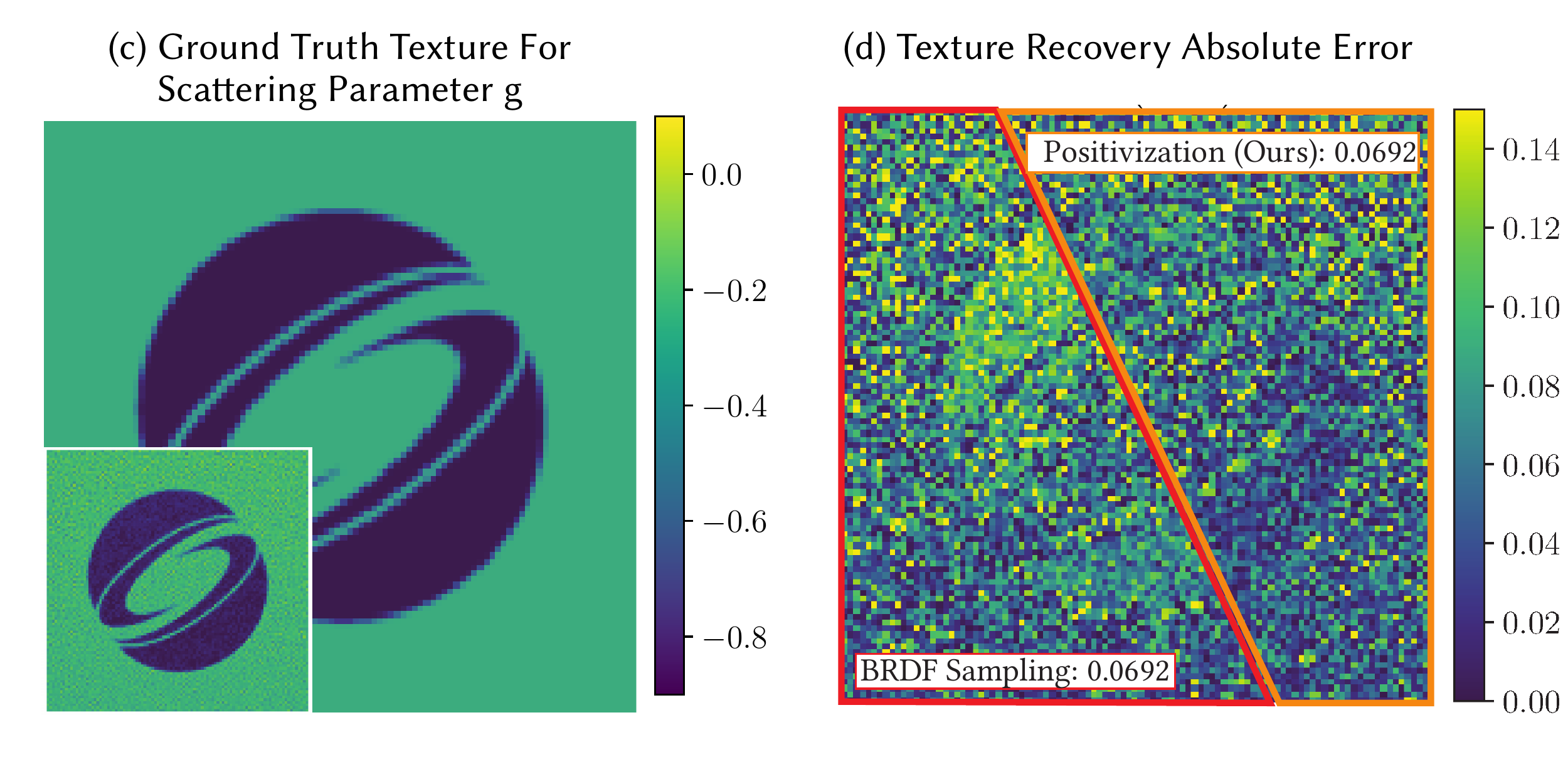}
    \caption{\textsc{Inverse Rendering of the scattering parameter $g$ of a Hanrahan-Krueger BRDF}.
    (a) forward rendering of the target.
    Our positivization has a lower texture recovery error than BRDF sampling (b).
    (c) shows the ground truth texture, with an inset of the recovered texture using positivization.
    (d) shows the error images for positivization and BRDF sampling.
    BRDF sampling is unable to recover the texture in the highly backscattering logo region. However, positivization can handle this region well.
    }
    \label{fig:invrend_pos}
    \vspace*{-0.15in}
\end{figure}

\begin{figure}[t]
    \includegraphics[width=\linewidth]{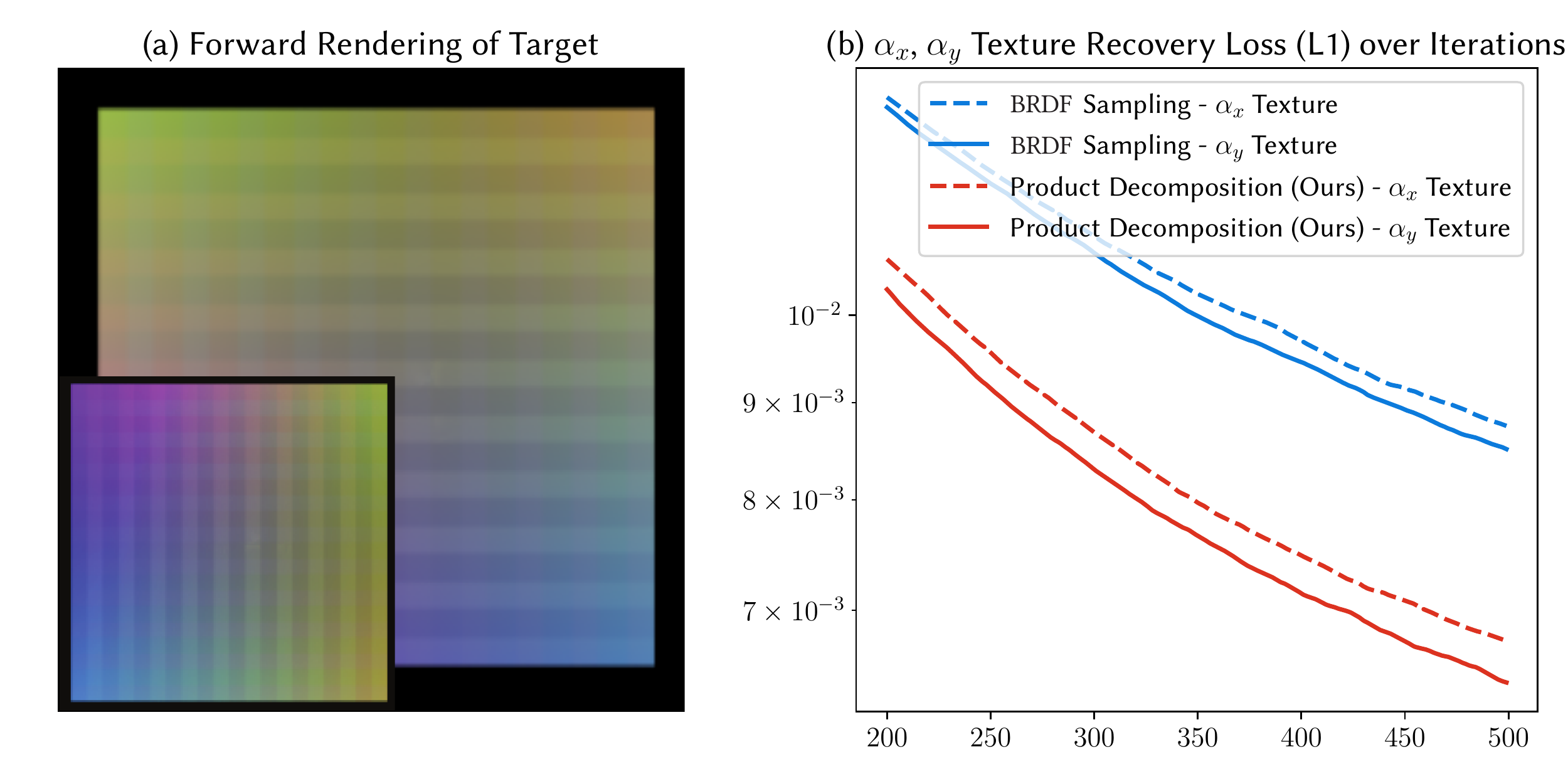}
    \includegraphics[width=\linewidth]{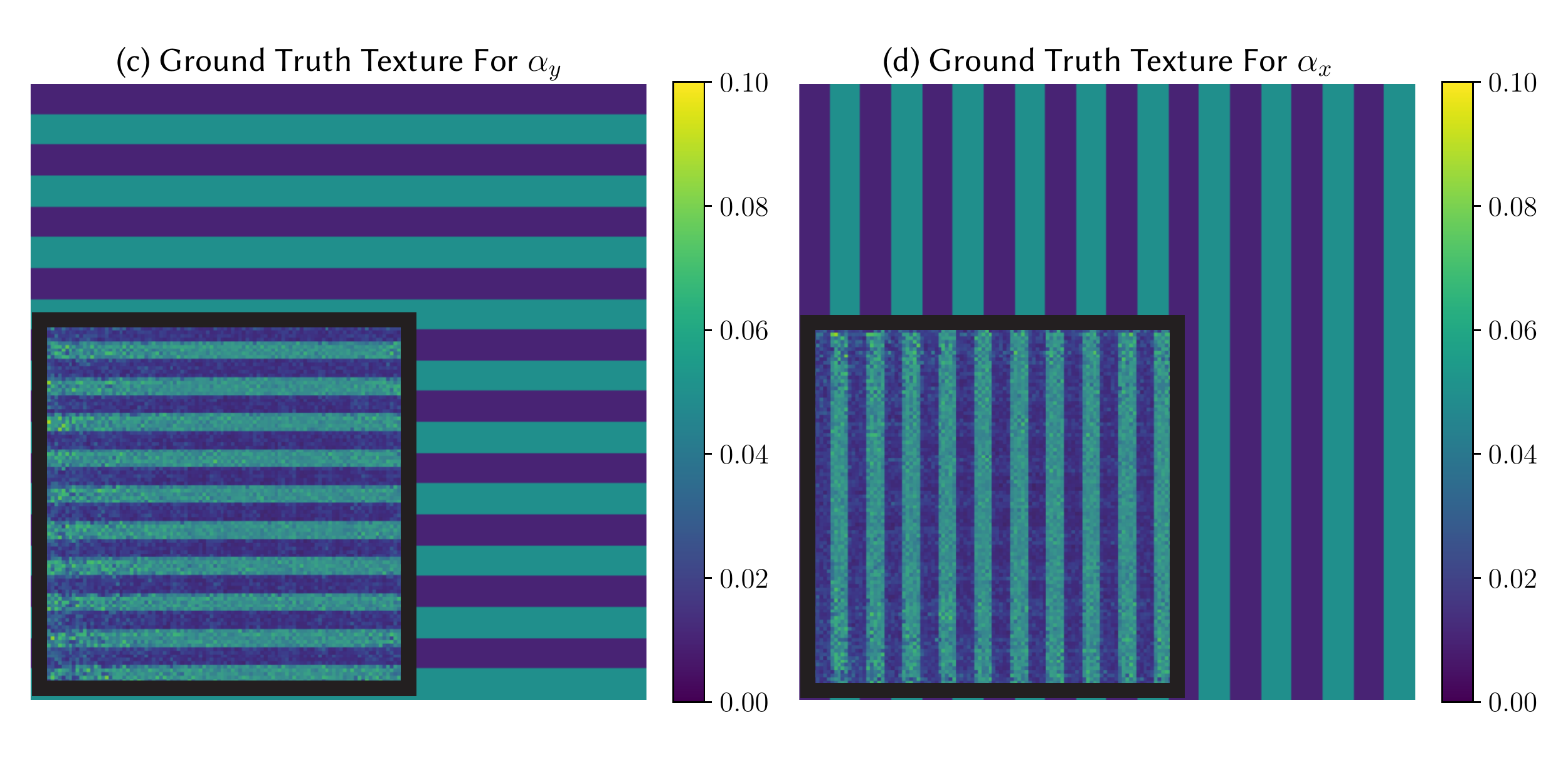}
    \caption{\textsc{Inverse Rendering of the roughness of an Anisotropic Beckmann Plate under a photometric stereo setup}.
    The forward rendering of the target under the two lighting setups is shown in (a) and its inset.
    The texture recovery loss (b) demonstrates that our product decomposition achieves lower texture recovery error than BRDF Sampling for both the $\alpha_x$ and $\alpha_y$ textures.
    (c) and (d) show the ground truth textures for $\alpha_x$ and $\alpha_y$, and the insets show our recovered textures.
    }
    \label{fig:invrend_prod}
    \vspace*{-0.15in}
\end{figure}

\begin{figure}[t]
    \includegraphics[width=\linewidth]{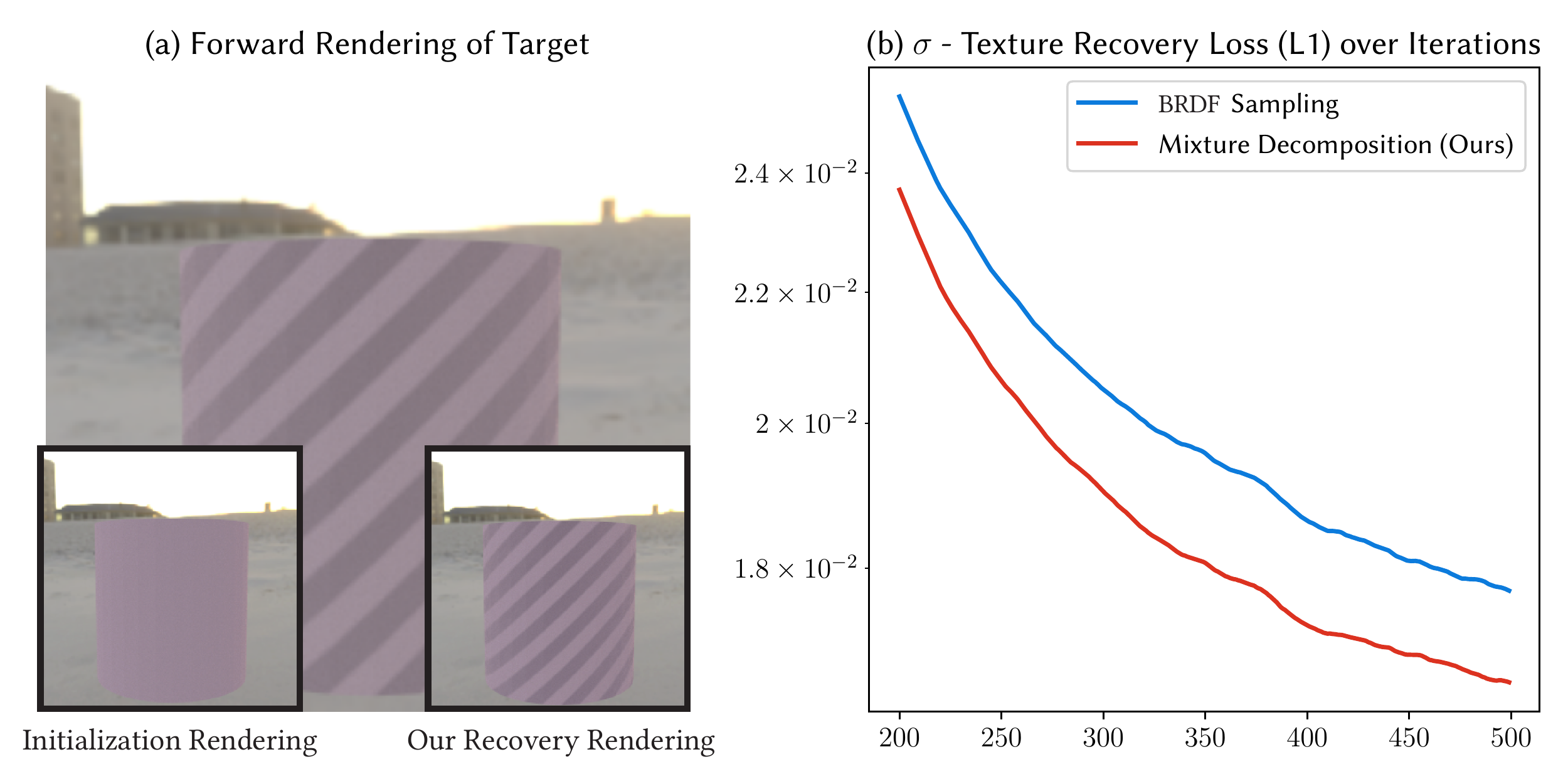}
    \caption{\textsc{Inverse Rendering of the roughness $\sigma$ of an Oren-Nayar BRDF}.
    (a) forward rendering of the target, insets show a rendering of the initialization and our final recovered texture.
    (b) Our mixture decomposition correctly deals with sign variance and has lower variance in gradients, and as a result has lower texture recovery error.
    }
    \label{fig:invrend_mixture}
    \vspace*{-0.15in}
\end{figure}

We demonstrate the benefits of correctly handling sign variance in gradients, for gradient-descent-based inverse rendering.
We apply inverse rendering to the task of spatially varying texture recovery, and evaluate the effectiveness of all three of our decompositions on it.
Our results for positivization are presented in \figref{fig:invrend_pos}, product decomposition in \figref{fig:invrend_prod}, and mixture decomposition in \figref{fig:invrend_mixture}.
All our inverse rendering results use 4 samples per pixel for both forward and gradient rendering at each optimization iteration.
We use the ADAM optimizer~\cite{Adam} and the respective loss graphs show the mean absolute texture recovery error ($L1$) after some initial iterations.

For positivization (\figref{fig:invrend_pos}), we recover the (spatially varying) scattering parameter $g$ of a Hanrahan-Krueger BRDF with the semi-infinite depth assumption, lit by a single area light.
The ground truth texture consists of a slightly back-scattering background region with $g = -0.3$, and a highly back-scattering logo region with $g = -0.9$, see \figref{fig:invrend_pos} (c).
The initialization is a back-scattering random initialization, i.e., $g < 0$. Positivization benefits from lowered gradient variance by correctly treating sign variance, and consistently has lower texture recovery error than BRDF sampling, see \figref{fig:invrend_pos} (b).
This is especially effective in the highly back-scattering logo region, where BRDF sampling suffers from high variance, whereas positivization which correctly accounts for sign variance can recover texture in this region better, see \figref{fig:invrend_pos} (d).

Positivization casts two shadow rays at each shading point. %
To ensure an equal ray-triangle intersection budget, we cast two shadow rays for BRDF sampling at each shading point.

For product decomposition (\figref{fig:invrend_prod}), we optimize the spatially varying anisotropic roughness textures ($\alpha_x$ and $\alpha_y$) of a Beckmann BRDF under a photometric stereo setup under two illumination conditions.
The two lighting conditions are rotated versions of the same environment map.
Starting from a random initialization for both textures, product decomposition's correct handling of the sign variance leads to a gradient estimator with lower overall variance, and consequently ensures lower texture recovery error across all iterations, as shown in \figref{fig:invrend_prod} (b).
The final recovery is displayed in \figref{fig:invrend_prod} (c),(d).

Our product decomposition computes the gradients for both roughness values using three samples at each shading point combined using multiple importance sampling (one each from $p_1$, $p_{2,x}$, $p_{2,y}$).
To ensure an equal-ray budget, we use three samples for BRDF sampling at each shading point too.

For mixture decomposition in \figref{fig:invrend_mixture}, we recover the spatially varying roughness of an Oren-Nayar BRDF under environment map illumination.
Once again, mixture decomposition benefits from lowered variance in gradients, and can recover a texture with lower error than BRDF sampling at an equal ray-triangle intersection budget, see \figref{fig:invrend_mixture} (b).

\section{Global Illumination}
\label{sec:gi}

\begin{figure}
\centering
\includegraphics[width=\linewidth]{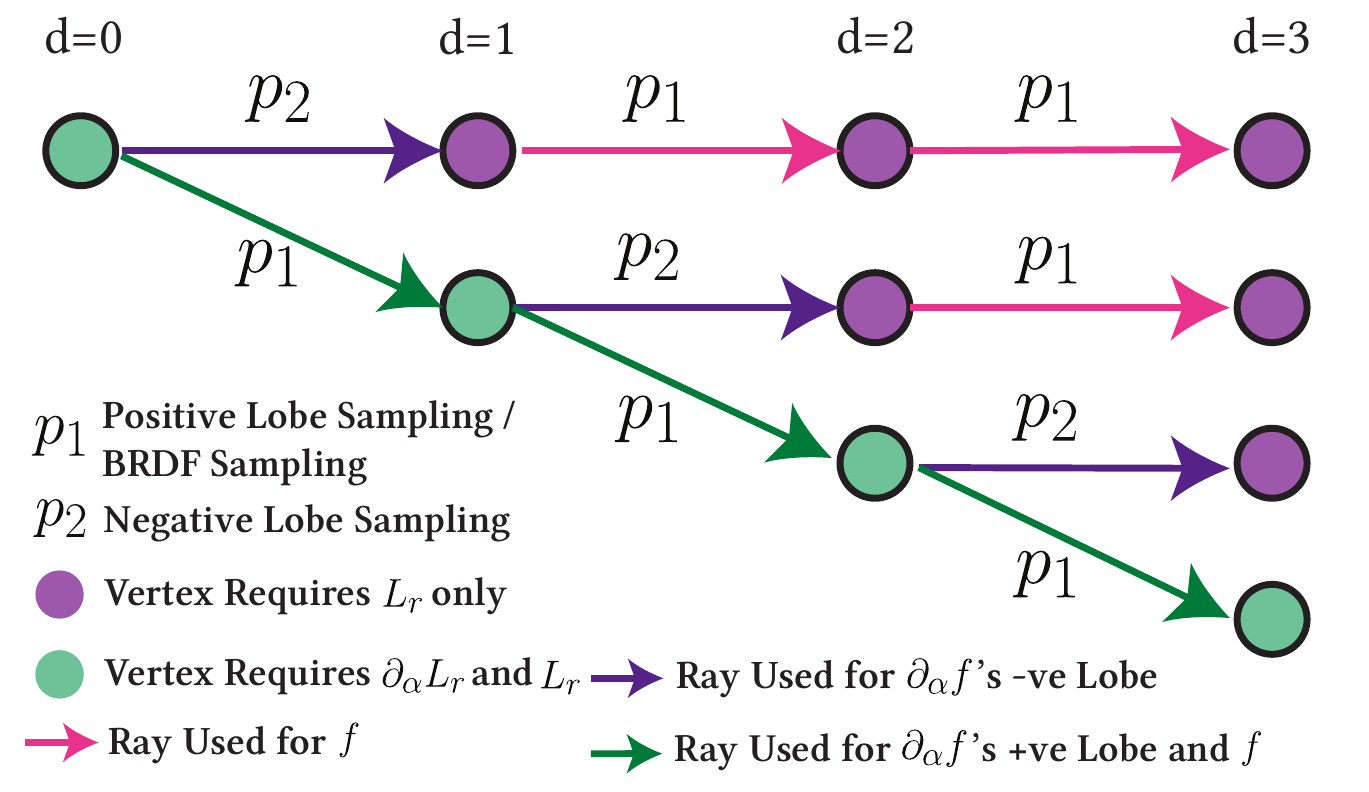}
\caption{
\textsc{Branching for Global Illumination with Max Depth=3 for Product \& Mixture Decomposition.}
Starting from a single vertex at $d=0$, estimating $\partial_\alpha L_r$ requires recursive estimation of both $L_r$ and $\partial_\alpha L_r$ via Eqns. \eqref{eq:d_rendering_gi_t1} and \eqref{eq:d_rendering_gi_t2}.
For product \& mixture decomposition, since one new branch is created at every vertex, the total number of rays used here is quadratic i.e $O(d^2)$, in the maximum depth $d$.
The sampling technique $p_1$ is the same for BRDF sampling and the positive lobe of the BRDF derivative, and $p_2$ is the sampling technique for the negative lobe.
For positivization, two new branches are created at each vertex corresponding to $p_+, p_-$, however, once created, they only require evaluation of $L_r$, and do not further branch out, so the complexity is still $O(d^2)$.
}
\label{fig:gi_branching}
\end{figure}

We now describe how to importance sample BRDF derivatives under multiple bounce global illumination.
The recursive rendering equation~\cite{kajiya} (ignoring emission) is given by a generalization of Eqn.~\eqref{eq:reflection},
\begin{align}
    L_r(\y, \wo; \alpha) = \int f(\y, \wi, \wo; \alpha) L_r(\z, -\wi; \alpha) \text{d}\wi,
    \label{eq:rendering_gi}
\end{align}
where we have substituted the incoming radiance $L_i(\y, \wi)$, with the outgoing/reflected radiance $L_r(\z, -\wi; \alpha)$, and $\z = \text{rayTrace}(\y, \wi)$ is the first intersection point from $\y$ in the direction $\wi$.
The recursive call of $L_r$ is a function of the BRDF parameter $\alpha$ because upon unrolling the recursion, it may be a function of an $\alpha$ dependent BRDF.
Differentiating this expression, we get,
\begin{align}
    \partial_\alpha L_r(\y, \wo; \alpha) &= \int \partial_\alpha f(\y, \wi, \wo; \alpha) L_r(\z, -\wi; \alpha) \text{d}\wi \label{eq:d_rendering_gi_t1}\\
    &+ \int f(\y, \wi, \wo; \alpha) \partial_\alpha L_r(\z, -\wi; \alpha) \text{d}\wi, \label{eq:d_rendering_gi_t2}
\end{align}
which recursively describes how differential radiance is reflected.
The two integrals (Eqn.~\eqref{eq:d_rendering_gi_t1} and ~\eqref{eq:d_rendering_gi_t2}) can be importance sampled separately.
We have seen how to importance sample Eqn. ~\eqref{eq:d_rendering_gi_t1} by applying different BRDF derivative decompositions in Sections ~\ref{sec:positivization}, ~\ref{sec:product_rule_decomposition}, and ~\ref{subsec:mixture_weights}.
Irrespective of the decomposition required, this requires two evaluations of $L_i$ corresponding to the positive and negative lobes and is done by regular path tracing (similar to the standard splitting approach~\cite{Arvo:1990:PTI}).
To importance sample Eqn. \eqref{eq:d_rendering_gi_t2}, we follow standard BRDF sampling and continue the same recursive importance sampling of $\partial_\alpha L_r$ at the next shading point.

This means that we need three samples at each shading point, one each for BRDF, positive lobe and negative lobe importance sampling.
Fortunately, for product and mixture decomposition, we can reduce this to two samples at each shading point.
For product decomposition, as we saw in \secref{sec:product_rule_decomposition}, one of either the positive or negative lobe decomposition PDFs is the same as BRDF sampling, and can share a sample with it.
For mixture decomposition, BRDF sampling can be simulated by randomly choosing a sample from either the positive or negative lobes with the probability equal to the mixture weight of the BRDF sampling strategy.

\begin{figure}[t]
    \includegraphics[width=\linewidth]{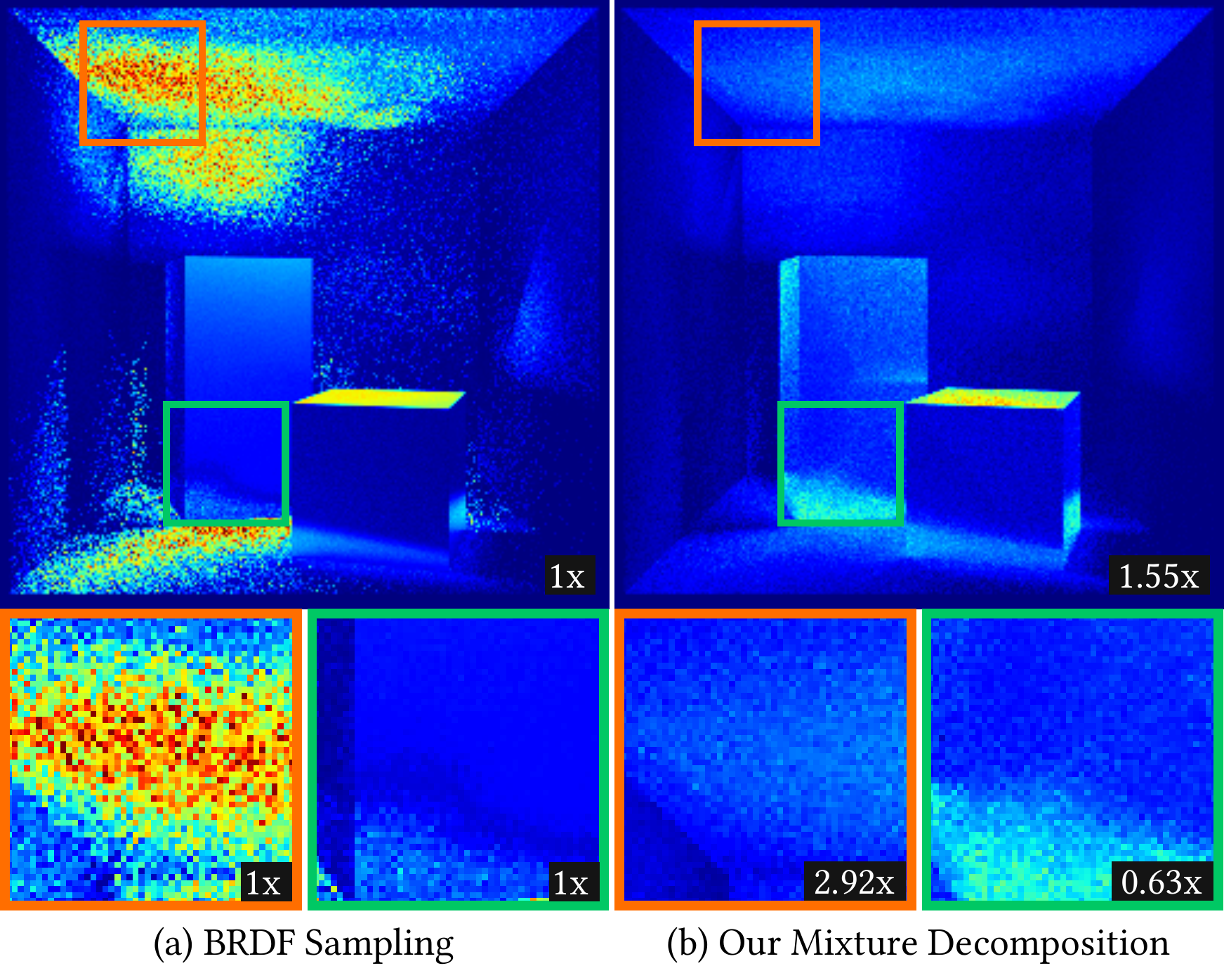}
      \vspace*{-0.5cm}
    \caption{
    Comparison between BRDF Sampling and Mixture Decomposition under 1 bounce Global Illumination and equal ray-triangle intersection budget.
    The overall BRDF is given by $f = w f_d + (1-w) f_s$ with $w=0.1$ for the left box and $w=0.9$ for the right box.
    We estimate the derivative with the weight $w$.
    $f_d$ is a lambertian diffuse lobe, and $f_s$ is an isotropic GGX lobe with $\alpha=0.05$.
    Mixture decomposition leads to lower variance in most regions since it correctly deals with sign variance.
    However, in some regions, e.g the green inset, the lighting, and visibility variance is more significant, and in that particular region, BRDF sampling happens to be better aligned to reduce this variance. 
    }
    \label{fig:gi_cbox}
    \vspace*{-0.25in}
\end{figure}

\vspace{0.1cm}
\noindent\textit{Branching Complexity and Comparison with BRDF sampling.}
Even though we use two samples to estimate Eqn.~\eqref{eq:d_rendering_gi_t1}, the total number of rays required to estimate $\partial_\alpha L_r$ for a maximum depth $d$ is quadratic i.e., $O(d^2)$, instead of exponential, see \figref{fig:gi_branching}, whereas it is $O(d)$ for BRDF sampling.
This is because we only apply splitting when estimating Eqn.~\eqref{eq:d_rendering_gi_t1}, which recurses on $L_r$, and we do not split when estimating Eqn.~\eqref{eq:d_rendering_gi_t2}.
The recursive call of $L_r$ in Eqn.~\eqref{eq:d_rendering_gi_t1} does not require splitting, which prevents exponential branching.
We have found that for one bounce of global illumination, with an equal-ray budget, our mixture decomposition can significantly reduce variance, see \figref{fig:gi_cbox}.

\section{Limitations and Future Work}

\noindent\textit{Determining the number of samples for each decomposed component.}
For all three decompositions, our current implementation applies a two-sample estimator which uses one sample per component.
It is possible that a different estimator can be more efficient in some cases.
For example, when the two components have different areas (i.e., $\int \partial_\alpha f_1 \neq \int \partial_\alpha f_2$ for components $f_1$ and $f_2$), it might be useful to adjust the number of samples according to the area of the component (we show in Appendix~\ref{sec:app_micro_int_zero} that microfacet normal distribution functions always have components with equal area).
Research in allocating budgets for multiple importance sampling can likely help in our case as well~\cite{he2014optimal,Sbert:2018:MIS,Grittmann:2022:EMI}.
Our estimator that always samples all components belongs to the \emph{deterministic mixture} scheme~\cite{OwenMCBook}.
An alternative is a \emph{random mixture}, which randomly chooses one component.
We opt for deterministic mixtures since they consistently outperform random mixtures in our direct lighting experiments (due to the stratification effect, similar to standard MIS v.s. \emph{one-sample} MIS).
For global illumination, random mixtures are the same as applying Russian Roulette to keep only one of the two branches, and can be more computationally convenient in some cases since they omit the need for quadratic branching.

\vspace{0.1cm}
\noindent\textit{Branching and Global Illumination}. %
Our adoption of deterministic mixtures requires path splitting for global illumination. 
While the branching complexity is quadratic instead of exponential (same as a bidirectional path tracer), it can add undesired overheads.
There are several ways to reduce the branching, 1) deterministically using only BRDF sampling or using random mixtures instead of deterministic mixtures after a certain recursion depth, 3) using path reconnection similar to Zhang et al.'s approach~\shortcite{Zhang:2020:PSDR}, to reconnect the branches back to a single primary path.

\vspace{0.1cm}
\noindent\textit{Better Optimization Schemes}.
Ultimately, for inverse rendering, the optimization is both ill-posed and non-convex. Recently, we have seen some work~\cite{rgbxy} which takes a step in this direction. We believe the study of efficient estimators of the derivatives is largely orthogonal and equally crucial.

\section{Conclusion}
Our importance sampling techniques provide a fundamental component for future differentiable rendering work, enabling correct handling of sign and shape variance of differential BRDFs.
BRDF sampling is widely used in forward rendering to deal with a variety of light transport phenomena; this includes unidirectional, bidirectional and gradient domain path tracing, Metropolis light transport, path guiding, photon mapping, etc.
Similarly, as the need to deal with the differentials of more complicated light transport phenomena arises, we will need to develop differential counterparts of these algorithms and we believe that our method will be well suited to serve as a fundamental building block for them.
Our product and mixture decompositions can also potentially have use outside of graphics for importance sampling real-valued functions.

\section*{Acknowledgements}
This research was supported by NSF Grants 2105806, 1703957, and the Ronald L. Graham Chair.
Our test scenes use 3D models from Turbosquid and HRDI environment maps from Polyhaven.
The Cornell Box scene was adapted from Mitsuba's example.

\bibliographystyle{ACM-Reference-Format}
\bibliography{sample-base}

\appendix

\section{BRDF Derivative Importance Sampling PDFs and CDFs}
All PDFs and CDFs are in solid angle coordinates, and do not include multiplication by $\sin\theta$ for change of variables to spherical coordinates.
PDFs may be defined in either $\wi$ or $\wh$ space, depending on the BRDF.
The PDFs defined in $\wh$ space must finally be transformed to $\wi$ space, and while doing so must include the appropriate jacobian $4\wo\cdot\wh$.
The PDFs are denoted by $p$ and their corresponding CDFs are $P$.
In the cases where CDFs are provided instead of inverse transform sampling routines, CDF inversion is done numerically.

\subsection{Positivization}
These are all isotropic BRDFs, and sampling for the azimuthal angle $\phi$ is uniform sampling.
We introduce PDFs and sampling routines for Blinn-Phong and Hanrahan-Krueger derivatives that have not been discussed in past literature to the best of our knowledge.

Importance sampling routines for the derivatives of isotropic GGX and Beckmann were first introduced by Zeltner et al.~\shortcite{Zeltner} in Appendix A of their paper, and we do not repeat them here.
However, they do not provide explicit formulae for the PDFs $p_+, p_-$ that we need for positivization.
These PDFs have a different normalization by a factor of $2$ than the PDF $p$ they use, so we define the PDFs $p_+, p_-$ here.

\subsubsection{Isotropic GGX}
\label{sec:app_iso_ggx}
\begin{align}
\begin{split}
r(\theta_h) &= \frac{8 \alpha^2 \sec^3\theta_h (\tan^2\theta_h - \alpha^2)}{ \left(\tan^2\theta_h  + \alpha^2 \right)^3
}\\
p_{\alpha,-}(\theta_h) &= -\min(r(\theta_h), 0)
\\
p_{\alpha,+}(\theta_h) &= 
 \max(r(\theta_h), 0)
\end{split}
\end{align}

\subsubsection{Isotropic Beckmann}
\label{sec:app_iso_beckmann}
\begin{align}
\begin{split}
 r(\theta_h) &= 
 \frac{4 e^{1 - tan^2\theta_h/\alpha^2} \sec^3\theta_h ( \tan^2\theta_h - \alpha^2)}{ \alpha^4
} \\
p_{\alpha,-}(\theta_h) &= -\min(r(\theta_h), 0)
\\
p_{\alpha,+}(\theta_h) &= 
 \max(r(\theta_h), 0)
\end{split}
\end{align}

\subsubsection{Blinn-Phong (Minnaert)}
\label{sec:app_blinn}

\begin{align}
\begin{split}
 r(\theta_h) &= 
 e(n+2)\cos^{n+1}\theta_h((n+2)\log\cos\theta_h + 1) \\
p_{n,-}(\theta_h) &= -\min(r(\theta_h), 0)
\\
p_{n,+}(\theta_h) &= 
 \max(r(\theta_h), 0) \\
P_{n,+}(\theta_h) &= -e(n+2)\cos^{n+2}\theta_h\log\cos\theta_h \\
P_{n,-}(\theta_h) &= 1-e(n+2)\cos^{n+2}\theta_h\log\cos\theta_h
\end{split}
\end{align}
For the Minnaert BRDF, the sampling routines are the same as above, but defined in $\theta_i$ space instead of $\theta_h$.

\subsubsection{Henyey-Greenstein (Hanrahan-Krueger)}
\label{sec:app_hk}

\begin{align}
\begin{split}
C &= \frac{3^{3/2}g^2(1-g^2)}{(3 + g^2)^{3/2} - 3^{3/2}(1-g^2)} \\
r(\theta_i) &= Cg^2
\frac{(g^2 + 3)\cos\theta_i + g(g^2 - 5)}
{
(g^2 - 2g\cos\theta_i + 1)^{5/2}
} \\
p_{g,-}(\theta_i) &= -\min(r(\theta_i), 0) \\
p_{g,+}(\theta_i) &= \max(r(\theta_i), 0) \\
P_{g,-}(\theta_i) &= \begin{cases}
    (1-C) - C\left[
 \frac{3g^2+1-g(g^2+3)\cos\theta_i}{(g^2 - 2g\cos\theta_i + 1)^{3/2}}
\right] ,& \text{if } p_{g,-}(\theta_i) > 0\\
    1,              & \text{otherwise}
\end{cases} \\
P_{g,+}(\theta_i) &= \begin{cases}
    C\left[
 \frac{3g^2+1-g(g^2+3)\cos\theta_i}{(g^2 - 2g\cos\theta_i + 1)^{3/2}}
\right] - 1 ,& \text{if } p_{g,+}(\theta_i) > 0\\
    1,              & \text{otherwise}
\end{cases}
\end{split}
\end{align}

\subsection{Product Decomposition}
For product decomposition, there are two sampling PDFs.
The first is $p_1 \propto g$, which is just regular BRDF sampling (e.g. visible normal distribution function sampling for GGX/ Beckmann); we do not repeat them here.
We provide importance sampling PDFs and CDFs for $\partial_\alpha g$.

For Anisotropic GGX and Beckmann, we provide the PDFs and importance sampling routines for $\partial_{\alpha_x}g$ with one of the directional parameters $\alpha_x$.
The corresponding PDFs and CDFs for the other directional parameter $\alpha_y$ can be obtained by swapping $\alpha_x$ with $\alpha_y$ and $\cos\phi_h$ with $\sin\phi_h$.
We do the same for Ashikhmin-Shirley too, except the directional parameters are $n_u, n_v$ in this case.

For the three BRDFs above, the CDF for $\phi_h$ generates an azimuthal angle in the range $[0, \pi/2]$.
$\phi_h$ is mirror symmetric about $\pi/2$ and has a period of $\pi$, which is used to transform $\phi_h$ to the range $[0, 2\pi]$ (and the jacobian needs to account for this via a division by $4$ as well).
The CDF for $\theta_h$ generates an elevation angle in $[0, \pi/2]$.

\subsubsection{Anisotropic GGX}
\label{sec:app_aniso_ggx}
Derivative with $\alpha_x$.
\\
\begin{equation}
\begin{split}
\label{eqn:pdf_ggx}
a(\phi_h) &= \frac{\cos^2\phi_h}{\alpha_x^2} + \frac{\sin^2\phi_h}{\alpha_y^2} \\
g(\theta_h, \phi_h) &=
\left( a(\phi_h)\sin^2\theta_h + \cos^2\theta_h \right)^{-2} \\
p_{\alpha_x}(\phi_h) &= \frac{4\cos^2\phi_h}{\pi\alpha_x^3\alpha_y a(\phi_h)^2} \\
 p_{\alpha_x}(\theta_h|\phi_h) &= \frac{4a(\phi_h)^2
 \tan^2\theta_h\sec^3\theta_h}{\left(\tan^2\theta_h a(\phi_h) + 1\right)^3} \\
P_{\alpha_x}(\phi_h) &= \frac{2}{\pi}\left[\ \tan^{-1}\left(\frac{\alpha_x}{\alpha_y}\tan\phi_h\right) \right. \\
&+ \left. \frac{\alpha_y \alpha_x \sin(2\phi_h)}{\alpha_x^2 + \alpha_y^2 + (\alpha_y^2 - \alpha_x^2)\cos(2\phi_h)}
\right] \\
P_{\alpha_x}(\theta_h|\phi_h) &= \frac{a(\phi_h)^2}{a(\phi_h)^2 - 1} \\
&- \frac{a(\phi_h)^2((1-a(\phi_h))\cos(4\theta_h) + a(\phi_h) + 3)}{4(a(\phi_h)^2 - 1)\left((a(\phi_h)-1)\sin^2\theta_h + 1\right)^2}
\end{split}
\end{equation}

\subsubsection{Anisotropic Beckmann (Ward)}
\label{sec:app_aniso_beckmann}
Derivative with $\alpha_x$.\\
The importance sampling PDFs and CDFs for the anisotropic Beckmann and Ward BRDFs are the same since the shape functions $g$ for both the BRDFs (and their derivatives) take on a similar functional form.
The PDF $p_{\alpha_x}(\phi_h)$ and CDF $P_{\alpha_x}(\phi_h)$ for them is the same as GGX, see Eqn. \eqref{eqn:pdf_ggx}.
Also see Eqn. \eqref{eqn:pdf_ggx} for the definition of $a(\phi_h)$.
\\
\begin{align}
\begin{split}
g(\theta_h, \phi_h) &= \sec^3\theta_h e^{-a(\phi_h)\tan^2\theta_h } \\
 p_{\alpha_x}(\theta_h|\phi_h) &= 2 a(\phi_h)^2 \tan^2\theta_h \sec^3\theta_h e^{-a(\phi_h) \tan^2\theta_h} \\
P_{\alpha_x}(\theta_h|\phi_h) &= 1 - (1 + a(\phi_h)\tan^2\theta_h) e^{-a(\phi_h)\tan^2\theta_h}
\end{split}
\end{align}

\subsubsection{Ashikhmin-Shirley}
\label{sec:app_ashk}
Derivative with $n_u$.
\begin{align}
\begin{split}
a(\phi_h) &= n_u\cos^2\phi_h + n_v\sin^2\phi_h \\
g(\theta_h, \phi_h) &= \cos\theta_h^{a(\phi_h)} \\
p_{n_u}(\phi_h) &= \frac{4 (n_u+1)^{3/2} \sqrt{n_v + 1}\cos^2\phi_h}{\pi (1 + a(\phi_h))^2} \\
 p_{n_u}(\theta_h|\phi_h) &= -\log\cos\theta_h(1 + a(\phi_h))^2\cos\theta_h^{a(\phi_h)} \\
P_{n_u}(\phi_h) &= \frac{2}{\pi} \left[\ \tan^{-1}\left(\sqrt{\frac{n_v + 1}{n_u + 1}}\tan\phi_h\right) \right. \\
&+ \left. \frac{\sqrt{(n_u+1)(n_v + 1)} \sin(2\phi_h)}{n_u + n_v + 2 + (n_u - n_v)\cos(2\phi_h)}
\right] \\
P_{n_u}(\theta_h|\phi_h) &= 1 - (1 - (a(\phi_h) + 1)\log\cos\theta_h) \cos\theta_h^{a(\phi_h) + 1}
\end{split}
\end{align}
\subsubsection{Microfacet ABC}
\label{sec:app_abc}

The ABC Microfacet BRDF is an isotropic microfacet BRDF, and so the sampling for $\phi_h$ is uniform.
The parameter $A$ does not play a role in the microfacet BRDF (it is canceled out by the normalization constant), so we ignore it, and only consider the derivatives with the parameters $B, C$.
\begin{align}
\begin{split}
g(\theta_h) &= (1 + B(1 - cos(\theta_h)))^{-C} \\
 p_{B}(\theta_h) &= \frac{B^2C(C-1)(B+1)^C(\cos\theta_h - 1) (1 + B(1 - \cos\theta_h))^{-1-C}}
 {(1 + BC - (B+1)^C)
 } \\
p_{C}(\theta_h) &= \frac{B(C-1)^2}{1 - (1+B)^{1-C}\left((C - 1)\log(B+1) + 1\right) }\frac{\log(1 + B(1 - \cos\theta_h))}{\left(1 + B(1 - \cos\theta_h)\right)^{C}} \\
 P_{B}(\theta_h) &= \frac{(B+1)^C \left( 1 + B(1 - \cos\theta_h) \right)^{-C} (1 + BC(1 - \cos\theta_h)) - (B+1)^C}
 {(1 + BC - (B+1)^C)
 } \\
 P_{C}(\theta_h) &= \frac{1 - (1 + B(1 - \cos\theta_h))^{1-C}((C-1)\log(1 + B(1 - \cos\theta_h)) + 1)}{1 - (B+1)^{1-C}((C-1)\log(1+B) + 1)}
\end{split}
\end{align}

\subsubsection{Hemi-EPD}
\label{sec:app_epd}
The Hemi-EPD microfacet BRDF is another isotropic BRDF, so $\phi_h$ is importance sampled using uniform sampling.
$\Gamma$ is the incomplete gamma function.

\begin{align}
\begin{split}
g(\theta_h) &= e^{\kappa\cos^\gamma\theta_h} - 1 \\
 p_{\kappa}(\theta_h) &= \frac{\gamma \kappa(-\kappa)^{1/\gamma}}{\Gamma(1 + 1/\gamma,0) - \Gamma(1 + 1/\gamma, -\kappa))}\cos^\gamma(\theta_h)e^{\kappa\cos^\gamma\theta_h} \\
 P_{\kappa}(\theta_h) &= \frac{\Gamma(1 + 1/\gamma, -\kappa\cos^\gamma\theta_h) - \Gamma(1 + 1/\gamma, -\kappa)}{\Gamma(1 + 1/\gamma,0) - \Gamma(1 + 1/\gamma, -\kappa))} \\
\end{split}
\end{align}

\subsubsection{Burley Diffuse BSSRDF}
\label{sec:app_bssrdf}
This BSSRDF is defined over an infinite plane, and is radially symmetric.
The polar angle $\phi$ is sampled uniformly.
We provide an importance sampling routine to sample the radial distance $r \in [0, \infty]$, for the derivative with the parameter $d$ that controls both its height and width.
Once again, a jacobian for multiplication with $r$ is required here.

\begin{align}
\begin{split}
g(r, d) &= \frac{e^{-r/d} + e^{-r/3d}}{r} \\
p_d(r) &= \frac{e^{-r/d} + e^{-r/3d}/3}{4d^2} \\
P_d(r) &= 1 - \frac{e^{-r/d}(r+d)}{4d} - \frac{e^{-r/3d}(3d + r)}{4d}
\end{split}
\end{align}

\subsection{Mixture Decomposition}
\subsubsection{Mixture Model}
\label{sec:app_mixture}
We are interested in differentiating a mixture model $f$, given by
\begin{equation}
\begin{split}
    f(\wi, \wo) &= w f_1(\wi, \wo) +
    (1 - w) f_2(\wi, \wo) \\
    \partial_{w} f(\wi, \wo) &= f_1(\wi, \wo) - f_2(\wi, \wo),
\end{split}
\end{equation}
with its parameter $w$.
Here, $f1$ and $f2$ are the two lobes of the BRDF.
The importance sampling scheme for the two terms of the derivative $\partial_w f$ are simply the BRDF importance sampling techniques for $f_1$ and $f_2$ respectively.

\subsubsection{Oren-Nayar}
\label{sec:app_on}
We are interested in differentiating the roughness $\sigma$.
The PDFs once again are in solid angle coordinates, not in spherical coordinates.
The first term of Eqn. \eqref{eq:on_deriv} requires standard cosine hemispherical sampling and we provide an importance sampling routine for the second term.
Here, $p_2(\theta_i)$ is made up of two terms depending on whether $\theta_i < \theta_o$, and they have weights $A_{21}^{'}, 1 - A_{21}^{'}$ respectively.
For $\phi_i$, an exact inverse transform sampling routine is available.
\begin{align}
\begin{split}
    A_{21} &= \frac{1}{2}\sin(\theta_o)(\theta_o - \sin(\theta_o)\cos(\theta_o))\\
    A_{22} &= \frac{1}{3}\tan(\theta_o)(1 - \sin^3(\theta_o))\\
    T_2 &= A_{21} + A_{22}, \; A_{21}^{'} = A_{21}/T_2, \\
    p_2(\theta_i) &= 
\begin{cases}
    A_{21}^{'} \frac{\sin(\theta_i)}{ (0.5(\theta_o - \sin(\theta_o)\cos(\theta_o)))} ,& \text{if } \theta_i < \theta_o\\
    (1 - A_{21}^{'}) \frac{3\sin(\theta_i) \cos(\theta_i)} {1 - \sin^3(\theta_o)},              & \text{otherwise}
\end{cases}\\
    p_2(\phi_i) &= 0.5 \max(0, \cos(\phi_o - \phi_i)) \\
P_2(\theta_i) &= 
\begin{cases}
    A_{21}^{'}\frac{\theta_i - sin(\theta_i)cos(\theta_i)}{\theta_o - sin(\theta_o)cos(\theta_o)} ,& \text{if } \theta_i < \theta_o\\
    A_{21}^{'} + (1 - A_{21}^{'})\frac{\sin^3(\theta_i) - sin^3(\theta_o)}{1.0 - sin^3(\theta_o)},              & \text{otherwise}
\end{cases} \\
\phi_i &= \begin{cases}
    \phi_o - sin^{-1}(2u) ,& \text{if } u < 0.5\\
    \phi_o + sin^{-1}(2u - 1),              & \text{otherwise}
\end{cases}
\end{split}
\end{align}

\subsubsection{Microcylinder}
\label{sec:app_microcylinder}
We want to importance sample the derivative of the BRDF corresponding to the volumetric scattering component $f_{r,v}$ in the original paper's notation, with the linear combination weight $k_d$. This BRDF does \textit{not} include cosine foreshortening. 
\begin{align}
\begin{split}
f(\wi, \wo) &= F\frac{(1 - k_d)g(\theta_h; \gamma_v) + k_d}{\cos\theta_i + \cos\theta_o}A \\
\partial_{k_d}f(\wi, \wo) &=  F\frac{1}{\cos\theta_i + \cos\theta_o}A - F\frac{g(\gamma_v, \theta_h)}{\cos\theta_i + \cos\theta_o}A,
\end{split}
\end{align}
where F is the Fresnel term, A is the albedo, and $g$ is a Gaussian with width $\gamma_v$.
The first term is importance sampled using cosine hemispherical sampling, which is also the importance sampling technique used for this BRDF in forward rendering.
The second term is importance sampled using inverse transform sampling for the Gaussian.

\section{Zeltner et al.'s Antithetic Sampling is a Special Case of Positivization}
\label{sec:zeltner_anti_pos}
Zeltner et al.'s~\shortcite{Zeltner} antithetic sampling involves generating \textit{paired and correlated} samples for the positive and negative lobes of the BRDF derivative $\partial_\alpha f$ in two separate passes, one pass for each lobe, and then \textit{averages} out the final result.

The correlation is induced by using the same random number generator state across the two passes.
The only difference between the two passes are that the first one uses a flag to trigger sampling from the positive lobe $p_+$ of the PDF $p = w p_+ + (1 - w) p_-$, and the second one triggers sampling from the negative lobe $p_-$.
Here, $w$ is the relative area of the positive lobe of $\partial_\alpha f$, given by $|\int \partial_\alpha f_+| / (|\int \partial_\alpha f_+| + |\int \partial_\alpha f_-|)$ and is equal to $0.5$ for the BRDF derivatives they consider, see Appendix \secref{sec:app_micro_int_zero}.

Their estimator for the integrand $\partial_\alpha f$ is given by,
\begin{align}
\begin{split}
    I = \frac{1}{2}\left(\frac{\partial_\alpha f(X_+)}{p(X_+)} + \frac{\partial_\alpha f(X_-)}{p(X_-)}\right),
    \label{eq:app_tiz_est}
\end{split}
\end{align}
where the samples are drawn from $X_+ \sim p_+$ and $X_- \sim p_-$, and the factor of $1/2$ comes from the fact that they average the result of the two passes.
We can further simplify Eqn. \eqref{eq:app_tiz_est}, to bring it in a form similar to the positivization estimator in Eqn. \eqref{eq:pos_est}, by noticing that $\partial_\alpha f(X) = \partial_\alpha f_+(X)$ when $X \sim p_+$ and similarly for $p_-$ too, which gives us
\begin{align}
\begin{split}
    I &= \frac{\partial_\alpha f_+(X_+(u))}{p_+(X_+(u))} + \frac{\partial_\alpha f_-(X_-(u))}{p_-(X_-(u))}.
\end{split}
\end{align}
The only difference between the estimator above and the positivization estimator is that the samples $X_-(u)$ and $X_+(u)$ are correlated because they use the same uniform random number $u$, whereas they are uncorrelated for positivization because positivization does not impose any such restriction.
Thus, antithetic sampling is a special case of \textit{positivization with correlated random numbers}.

Positivization (with uncorrelated random numbers) achieves its variance reduction due to the stratification of the real-valued function into positive and negative functions, and we have experimentally verified that antithetic sampling (with correlated random numbers) consistently has similar variance reduction as positivization.
As a result, antithetic sampling's variance reduction can be explained by the \textit{implicit stratification of $\partial_\alpha f$ into positive and negative lobes}.
See \figref{fig:res_zeltner_chang} for an example of the variance reduction.

\section{Microfacet BRDF Derivatives Integrate to Zero}
\label{sec:app_micro_int_zero}
Previous work~\cite{Zeltner} has noted that the derivative of the normal distribution function of the isotropic GGX (and Beckmann) BRDF with its roughness parameter has positive and negative lobes with equal area.
Here, we prove that this observation extends to \textit{all the derivatives of all microfacet normal distribution functions}.

The projected area of a microfacet BRDF's normal distribution function $D$ always integrates to $1$ i.e a constant,
\begin{align}
\int D(\wh, \alpha) \cos\theta_h \text{d}\wh = 1
\end{align}
As a result, its derivative with any parameter $\alpha$ integrates to $0$,
\begin{align}
\int \partial_\alpha D(\wh, \alpha) \cos\theta_h \text{d}\wh = 0
\end{align}
which means that the \textit{positive and negative lobes of $\partial_\alpha D \cos\theta_h$ have equal area}.
Since we generally construct microfacet derivative sampling PDFs proportional to the derivative of the projected normal distribution function, the sampling PDFs (irrespective of the decomposition) for the positive and negative lobes of the derivative must \textit{necessarily} have equal area.

\end{document}